\documentclass[12pt]{article}%

\usepackage[latin1]{inputenc}
\usepackage[T1]{fontenc}
\usepackage{amsfonts,amssymb,amsmath, epsfig}
\usepackage{color,graphicx,graphics}
\usepackage{amsmath,amstext,amssymb,amsfonts, amscd}
\usepackage{hyperref}
\usepackage{a4wide}
\usepackage{booktabs}
\usepackage{cite}

\usepackage{cleveref}
\numberwithin{equation}{section}
\usepackage[labelformat=simple]{subcaption}

\renewcommand{\u}{\tilde{u}}

\def\Box{\leavevmode\vbox{\hrule
		\hbox{\vrule\kern4pt\vbox{\kern4pt}%
			\vrule}\hrule}}

\newtheorem{lemma}{Lemma}[section]

\newtheorem{definition}[lemma]{Definition}

\newtheorem{remark}{Remark}[section]
\newtheorem{example}{Example}[section]

\def\softd{{\leavevmode\setbox1=\hbox{d}%
          \hbox to 1.05\wd1{d\kern-0.4ex{\char039}\hss}}}

\definecolor{Cgrey}{rgb}{0.85,0.85,0.85}

\newcommand{\cmag}{\color{magenta}}

\newcommand{\Ov}[1]{\overline{#1}}

\newcommand{\mesh}{\mathcal{T}}
\newcommand{\TS}{\Delta t}
\newcommand{\Div}{{\rm div}_x}
\newcommand{\Grad}{\nabla_x}

\newcommand{\Lap}{\Delta}

\newcommand{\vr}{\varrho}
\newcommand{\vm}{\mathbf{m}}

\usepackage{adjustbox}
\usepackage{lipsum}
\usepackage{multirow}

\begin {document}
\title{Quantum algorithms for Young measures -- applications to nonlinear partial differential equations}

\author{Shi Jin\footnote{School of Mathematical Sciences, Institute of Natural Sciences,
Ministry of Education Key Laboratory in Scientific and Engineering Computing, Shanghai Jiao Tong University, Shanghai 200240, China, shijin-m@sjtu.edu.cn.}, \
Nana Liu\footnote{Institute of Natural Sciences, Global College,
Ministry of Education Key Laboratory in Scientific and Engineering Computing, 
 Shanghai Jiao Tong University, Shanghai 200240, China, {\tt nana.liu@quantumlah.org}}, \  M\'aria Luk\'a\v{c}ov\'a-Medvi{\softd}ov\'a\footnote{Johannes Gutenberg-University Mainz, Institute of Mathematics, Staudingerweg 9, 55128 Mainz, Germany, \\
RMU Co-Affiliate TU Darmstadt, Germany {\tt lukacova@uni-mainz.de}} \ and\
Yuhuan Yuan\footnote{School of Mathematics, Nanjing University of Aeronautics and Astronautics, 
Jiangjun Avenue No. 29, 211106 Nanjing, P. R. China,
{\tt yuhuanyuan@nuaa.edu.cn}}
~~ }

\date{\today}
\maketitle
\begin{abstract}
 Many nonlinear PDEs have singular or oscillatory solutions or may exhibit physical instabilities or uncertainties. This requires a suitable concept of physically relevant generalized solutions. Dissipative measure-valued solutions have been an effective analytical tool to characterize PDE behavior in such singular regimes. They have also been used to characterize limits of standard numerical schemes on classical computers.  The measure-valued formulation of a nonlinear PDE  yields an optimization problem with a linear cost functional and linear constraints, which can be formulated as a linear programming problem. However, this linear programming problem can suffer from  the curse of dimensionality. In this article, we propose solving it using quantum linear programming (QLP) algorithms and discuss whether this approach can reduce costs compared to classical algorithms.   
 We show that some QLP algorithms, such as the quantum central path algorithm, have up to {\it polynomial} advantage over the classical interior point method.  
 For problems where one is interested in the dissipative weak solution, namely the expected values of the Young measure, we show that the QLP algorithms offer no advantage over direct classical solvers. Moreover, for random PDEs, there can be up to  {\it polynomial} advantage in obtaining the Young measure over {\it direct} classical PDE solvers. This is a significant advantage  over standard  PDE solvers, since the  Young measure 
 provides a more detailed description of the solution.
 We also propose some open questions for future development in this direction. 

\end{abstract}

\noindent {\bf Keywords.}  quantum algorithms, linear programming, nonlinear partial differential equations, uncertainty quantification, random parameterized Young measures, moment closure problems 

\medskip

\noindent {\bf AMS Subject Classification. } 60C50, 60H35, 82C40

\tableofcontents 

\section {Introduction}

 The development of quantum algorithms for scientific computing problems has quickly emerged as an active area of research due to their potential for providing up to polynomial or even exponential quantum advantage, subject to certain caveats. The first milestone in this direction is the HHL algorithm for solving linear systems of equations \cite{HHL2009}, which achieves an up to exponential advantage in the quantum linear systems of equations problem.  PDEs pose tremendous hurdles for classical simulation when the dimension is high, like in quantum dynamics and kinetic theory; or when small or multiple scales require high-resolution numerical simulation, like in fluid dynamics, turbulence, and other multiscale problems. Due to the potential in using quantum simulation to alleviate the curse of dimensionality, the development of efficient quantum algorithms is an attractive direction. There are several approaches in quantum algorithms for linear PDEs. The first one fully discretizes the PDEs, making it a linear algebra problem; then HHL-like quantum linear algebra solvers are used \cite{BerryChilds2017ODE}. The second uses {\it Schr\"odingerization} \cite{schrpra,schrprl, 2023analogPDE} that maps a non-unitary evolution into a unitary one by adding one higher dimension, where unitary evolution is most suitable for quantum simulation.  This is a quite general framework for all linear dynamical systems, and is suitable for both qubit-based  (see the related approach LCHS \cite{ALL2023LCH}) as well as continuous variable (qumodes) \cite{2023analogPDE} platform, making it competitive for both general and analog quantum computing.

 The nonlinear PDEs, on the other hand, have remained one of the main challenges for quantum simulation, and so far, there has been limited success in developing quantum algorithms that can demonstrate quantum advantage for truly nonlinear regimes. Since quantum computation is based primarily on linear processes, it becomes necessary to {\it linearize} nonlinear PDEs to make them suitable for quantum computation. The first such approach uses the Carleman truncation, whose validity is limited to weak nonlinearity and strong dissipation \cite{Liu-PNAS}. For most physical problems, like those studied in this article, the nonlinearity is strong and dissipation weak. This regime is challenging not only because of the nonlinearities, but also due to the singular behavior brought in by the nonlinearity, such as shocks, caustics, blow-ups, ill-posedness, etc.\ that cannot be easily handled by naive linearizations that do not yield the physically relevant {\it weak or generalized} solutions.  

 One of the most important nonlinear PDEs that has strong nonlinearity and weak dissipation is the Hamilton-Jacobi equation, where efficient quantum algorithms have been developed for both multivalued solutions \cite{JinLiu2022nonlinear} and viscosity solutions \cite{hjviscosityarxiv}-- both of which are of physical interest but are relevant in different applications-- that are valid beyond the formation of caustics. For multi-valued solutions, the level set method is used to map the equation {\it exactly} into the phase space {\it linear} Liouville equation, which can be handled by quantum simulation. For viscosity solutions,  an entropy-penalization method was used to approximate the Hamilton-Jacobi equations by a  discrete time nonlinear dynamical system which, via the Cole-Hopf transformation, gives rise to a heat equation like linear dynamics that can be solved by quantum algorithms.

\paragraph{The goal of this paper}  is to launch the program of developing \emph{Young measure-based quantum algorithms} for broader classes of nonlinear PDEs.  
The notion of Young measures was introduced as a rigorous analytical framework for nonlinear PDEs that exhibit singular and/or oscillatory and concentration behavior, in physical systems such as nonlinear hyperbolic systems and gas dynamics, homogenization problems in composite materials, Hamilton-Jacobi equations in optimal control, etc.,  in which the characterization of weak limit of the solution becomes necessary but difficult \cite{MR3409135,MR257325}, see also \cite{DiPerna1985}.  
Instead of solving nonlinear PDEs in the class of distributional (weak) solutions, we extend the solution class to measure-valued solutions.
The latter is represented by the space-time parametrized probability measure, the so-called Young measure. 

While the concept of  Young measure was introduced as a theoretical tool in modern mathematical analysis and nonlinear PDEs, its use as a novel computational tool has seen increasing activity only in recent years \cite{Chertock2026, chu2025solving, MR4390192, fjordholm2016computation, cardoen2024}.
However, computing the Young measure directly suffers from the curse of dimensionality for classical computations. A  main feature of the measure-valued formulation is that 
it transforms a nonlinear system of PDEs into a {\it linear} programming problem with {\it linear} constraints. Thus, it provides a natural setup for quantum computation, whose evolution is intrinsically linear and naturally handles linear dynamics.

\paragraph{Dissipative measure-valued solutions and computation of the Young measures}
 Our approach is motivated by recent works on 
 \emph{dissipative measure-valued solutions} to compressible fluid flows; see, e.g., \cite{MR4390192,MR4130543,MR4674068}, where convergence of classical numerical algorithms was analyzed for relevant problems. Dissipative measure-valued solutions are a weak closure of consistent and stable approximations  \cite{FL_18, FLSS_22, MR4390192,MR4130543,MR4674068}. Thus, they naturally arise as weak limits of numerical approximations.  Dissipative measure-valued solutions provide a framework for capturing oscillations and concentrations, which commonly arise in phenomena such as turbulent flows. Measure-valued formulation thus allows to connecting large-scale fluid behavior with unresolved microscopic dynamics, offering a coherent description across different scales.

 Although the class of
 dissipative measure-valued solutions of fluid flows  is larger than the class of classical or weak solutions, it enjoys several important properties.
 If a strong solution exists, then all dissipative measure-valued solutions coincide with the classical (smooth) solution emanating from the same initial data. 

 For a given initial data, all dissipative measure-valued solutions create a convex, bounded, and compact set in a suitable topology. Thus, dissipative measure-valued solutions are well suited to the variational formulation of selection criteria for determining a unique solution or selecting a physically desirable solution. For example, for the compressible Euler equations, which are known to be non-unique in the class of weak entropy solutions, selection criteria based on the entropy and energy defect maximization in the class of dissipative measure-valued solutions were proposed to recover well-posedness of the Euler equations, cf.~\cite{Feireisl-Jungel-Lukacova, Feireisl-Lukacova}.

 \paragraph{Novelties of our approach}

 In the present work, we use a measure-valued formulation {\it directly} and construct \emph{Linear Programming (LP)} algorithms to approximate the corresponding density distribution function. Due to the high dimensionality of this function, we use \emph{ quantum algorithms} to reduce the cost associated with dimensionality. In a digital quantum algorithm, one needs only $n$ qubits to simulate a $ 2^n$-dimensional space.   
 
 The rationale for our program is as follows.  First, quasilinear PDEs such as the Euler and Navier-Stokes equations may have non-unique weak (distributional) solutions \cite{Chiodaroli-DeLellis-Kreml:2015, DeLellis-Szekelyhidi:2009}, or instability (such as the Rayleigh-Taylor instability in fluid dynamics), and Young measure helps to determine or characterize the physically relevant unique solutions in several scenarios \cite{DiPerna:1983a,DiPerna-Majda:1987} that could not be obtained or understood directly by solving the original PDEs. It was pointed out by Glimm etc. in \cite{Lim-Glimm} that "{\it weak solutions in the form of a space
time dependent pdf for the primitive variables are actually what is required scientifically for
a combustion simulation."} 

Secondly, these physical systems may contain uncertainties in initial and/or boundary data, coefficients in the equations, etc., that call for uncertainty quantification (UQ). Usually, uncertainties lie in high-dimensional dimensions, and Young measure is a convenient tool to conduct UQ or moment closures for these problems \cite{cardoen2024, chu2025solving}. 

Thirdly, there are many physical or engineering problems in which one is not interested in solutions of PDEs at each point, but rather their probability measure, in order to understand, for example, the maximum values of certain quantities of interest 
(maximum lift on a wing,  peak temperature in a reactor, optimal bounds on dissipation rate in turbulent flows without tracking every individual eddy) \cite{Lim-Glimm}, etc. Therefore, developing efficient algorithms to compute the probability measure is of significant interest in its own right. Finally, the Young measure formulation transforms the original nonlinear PDEs into a {\it Linear Programming}  problem in which both the objective functions and constraints are {\it all linear}, and thus, it sets the foundation for possible development of efficient quantum algorithms for many nonlinear PDEs.

We show the performance of several quantum linear programming (QLP) algorithms to solve the LP problems for Young measures associated with nonlinear PDEs, and compare their complexities  or costs with the classical LP solvers. We find that certain quantum LP solvers like the quantum central path (QCP) algorithms provide {\it algebraic} advantages over the classical LP algorithms, for both deterministic and random nonlinear PDEs,  in obtaining the Young measure. For random PDEs (not deterministic PDEs), the quantum cost for obtaining the Young measure can be up to {\it polynomially} more efficient than even direct classical PDE solvers, where the latter provides much less information than the Young measures themselves. These results make the quantum algorithms for computing the Young measures--not the solution of the original PDEs--attractive. 

Nevertheless, 
we have not demonstrated quantum advantage when compared with the classical solvers directly solving the original nonlinear PDEs. 
To achieve quantum advantages,   more efficient QLP algorithms need to be developed.  This is an important future direction of research. 

\paragraph{Organization of the paper}
The paper is organized as follows. In Section~\ref{sec:problemsetting}, we introduce the concept of the measure-valued solutions and the corresponding Young measures for nonlinear PDEs and show how they can be computed via LP problems. To demonstrate the  potential of our approach for broad classes of nonlinear PDEs, we consider three prototype nonlinear PDEs:
the inviscid Burgers equation, the barotropic Euler equations, and the Allen-Cahn equation, and present numerical results for the resulting LP problems using classical computers, see Appendix~\ref{app:simulations}. In Section~\ref{sec:youngnonlinearpde1} we introduce quantum algorithms for the corresponding LP problem for deterministic nonlinear PDEs, and compare their performance to classical LP algorithms, and compute costs for the three prototype nonlinear PDEs. In Section~\ref{sec:uncertain}, we compare the costs of classical and quantum algorithms for nonlinear PDEs with uncertain parameters, and conclude with an outlook and open questions in Section~\ref{sec:outlook}.

\section{Problem Setting} \label{sec:problemsetting}

To explain the ideas of the measure-valued solutions and the construction of the LP method, let us start with a nonlinear hyperbolic conservation law

\begin{eqnarray}
\label{PDE1}
&& \partial_t u(t,x) + \Div f(u(t,x))  = 0,   \qquad   (t,x) \in (0,T)\times \Omega, \quad \Omega \subset \Bbb{R}^d,
\end{eqnarray}
subject to the initial condition $u(0,x)=u_0(x)$ and suitable  boundary condition, e.g.~periodic boundary conditions. 

Here, $u(t,x)\in \Bbb{R}^n$ is the conservative variable and $f\in \Bbb{R}^{n\times d}$ is the corresponding flux. 
Furthermore, the Second Law of Thermodynamics selects the solution satisfying the entropy inequality
\begin{align}
\label{PDE2}
\partial_t \eta(u(t,x)) + \Div q(u(t,x))  \geq  0,
\end{align}
where $\eta \in \Bbb{R}$ is the total physical entropy and $q \in \Bbb{R}^d$ is the corresponding entropy flux.  Note that the entropy function $\eta$ is a concave function of $u$.

\medskip
We start by introducing the weak entropy  solution of \eqref{PDE1}:
\begin{definition}
 $u(t,x)$ is a weak solution to \eqref{PDE1} if 
\begin{align}\label{weak-soln}
\int_{0}^T \int_{\Omega} \bigg( \partial_t \psi(t,x) \cdot u(t,x)  + \Grad  \psi(t,x) :  f(u(t,x)) \;  \bigg) d x d t  = 0,
\end{align}
for {all test functions} $\psi \in C^1_c((0,T) \times \Omega; \Bbb{R}^n)$. It satisfies the entropy condition if
\begin{align} \label{weak-entropy}
\int_{0}^T \int_{\Omega} \left( \partial_t \tilde{\psi}(t,x)  u(t,x)   + \Grad  \tilde{\psi}(t,x) \cdot q(u(t,x)) \; \right) d x d t  \leq 0,
\end{align}
for {all test functions} $\tilde{\psi} \in C^1_c((0,T) \times \Omega)$, $\tilde{\psi} \geq 0$.
\end{definition}
\noindent Here, we have used the symbol "$:$" for  
the Hadamard product, which is an entry-wise multiplication of two matrices.

For theoretical results on existence and uniqueness of weak entropy solutions for one-dimensional systems of hyperbolic conservation laws and multidimensional scalar hyperbolic conservation laws, we refer to \cite{cardoen2024, MR1816648,MR3468916,MR3793404}. 

The situation changes dramatically for multidimensional systems of hyperbolic conservation laws. As shown in the pioneering works of 
De Lellis and Sz\'ekelyhidi~\cite{DLSz, DlSz1}, there might exist infinitely many weak entropy solutions, see also \cite{Klingenberg1, CF_21} and the references therein. As shown recently in \cite{CF}, the set of initial data for which infinitely many weak entropy solutions of the compressible Euler equations exist is dense in $L^p$-topology. We also refer to recent works, where various selection criteria were studied to recover well-posedness of the compressible Euler equations in the class of measure-valued or dissipative (measure-valued) solutions \cite{Breit, klingenberg2025, Feireisl-Jungel-Lukacova, feireisl-lukacova2025}.

Young measure is an elegant tool for characterizing weak limits of oscillating and concentrating sequences and has been shown to be particularly well-suited to describe low-regularity solutions of nonlinear PDEs. In addition, it is a convenient tool to analyze uncertainty quantification in many problems arising in  physics, biology, or engineering.    
Motivated by the selection criteria proposed in \cite{Feireisl-Jungel-Lukacova, 
Feireisl-Lukacova, feireisl-lukacova2025} and by \cite{chu2025solving},  we seek 
 a direct approximation of the Young measure that generates a measure-valued solution of \eqref{PDE1}-\eqref{PDE2}, and highlight the potential of this approach for QLP.

\subsection{Measure-valued solution}

Next we  introduce the concept of measure-valued solution of \eqref{PDE1}-\eqref{PDE2}.

\begin{definition}
The Young measure $\mathcal{V}_{t,x} \in \mathcal{P}(\mathbb{R}^n)$ with 
\begin{align}\label{initial-ym-0}
\mathcal{V}_{0,x}(\cdot) = \delta(\cdot - u_0(x) ) \mbox{ for  a.a. } x  \in  \Omega
\end{align}
 is a measure-valued solution of \eqref{PDE1}-\eqref{PDE2} if 
\begin{align}\label{PDE-YM}
\int_{0}^T \int_{\Omega} \left( \partial_t \psi(t,x) \cdot \int_{\mathbb{R}^n} \u \;  \mathrm{d}\mathcal{V}_{t,x}(\u)   + \Grad  \psi(t,x) : \int_{\mathbb{R}^n} f(\u) \; \mathrm{d}\mathcal{V}_{t,x}(\u) \right) d x d t  = 0,
\end{align}
for {all test functions} $\psi \in C^1_c((0,T) \times \Omega; \Bbb{R}^n)$;
\begin{align} \label{PDE-YM-entropy}
\int_{0}^T \int_{\Omega} \left( \partial_t \tilde{\psi}(t,x)  \int_{\mathbb{R}^n} \eta(\u) \;  \mathrm{d}\mathcal{V}_{t,x}(\u)   + \Grad  \tilde{\psi}(t,x) \cdot \int_{\mathbb{R}^n} q(\u) \; \mathrm{d}\mathcal{V}_{t,x}(\u) \right) d x d t  \leq 0,
\end{align}
for {all test functions} $\tilde{\psi} \in C^1_c((0,T) \times \Omega)$, $\tilde{\psi} \geq 0.$   
\end{definition}

For simplicity of presentation, we do not consider here a concentration measure that may arise in an unbounded case. Existence of measure-valued solutions, their compatibility, and measure-valued--strong uniqueness principle for multidimensional hyperbolic conservation laws \eqref{PDE1}-\eqref{PDE2}  were investigated, e.g., in \cite{MR4390192}. 
Moreover, provided that $\mathcal{V}_{t,x}$ is a Dirac measure concentrated at $u=u(t,x)$ {for every $(t,x)\in[0,T]\times \Omega$},  the standard weak formulation of \eqref{PDE1}-\eqref{PDE2} is recovered. 

\subsection{Linear programming problem}\label{sec:LP-Determin}

Without loss of generality, we assume that the Young measure $\mathcal{V}_{t,x}$  has at each $(t,x)$ the corresponding probability distribution function $F(t,x,\cdot)$, that is an absolutely continuous random variable on $\mathbb{R}^n$ with respect to $\mathcal{P}(\mathbb{R}^n)$. 
Let us denote it with
\begin{align*}
& F: (0,T)  \times \Omega \times \mathbb{R}^n  \to [0,\infty), \\
& \mbox{satisfying} \quad  \int_{\mathbb{R}^n} F(t,x,\xi) d\xi=1, \qquad F(t, x, \xi) \geq 0, \qquad \xi \in \mathbb{R}^n,  
\end{align*}
which represents the probability that $u(t,x)$ attains a value $\xi$.

Due to the linearity of \eqref{PDE-YM}-\eqref{PDE-YM-entropy}, there may be many Young measures satisfying \eqref{PDE-YM}-\eqref{PDE-YM-entropy}. 
On the other hand, the entropy inequality \eqref{PDE2} or \eqref{PDE-YM-entropy} is used to single out unphysical solutions, whereas the election criterion (entropy maximization and/or energy minimization) is used to obtain the unique solution. Both conditions reflect the Second Law of Thermodynamics; the selection criterion needs to be applied in addition to the entropy inequality if the latter does not yield a unique solution.

Consequently, we formulate \eqref{PDE1}-\eqref{PDE2} as a linear optimization problem with the cost functional realizing the entropy maximization:
\begin{align}\label{LP1}
    \text{argmax}_{F(t, x, \xi)} \int_{\Omega} \int_{\mathbb{R}^n} \eta(\xi) F(t, x, \xi) d \xi dx
\end{align}
subject to:
\begin{align}\label{LP2}
& \int_{\mathbb{R}^n} F(t,x,\xi) d\xi=1,   \qquad  F(t, x, \xi) \geq 0,\\
         \label{LP3}
&  \int_{\mathbb{R}^n} \left( \xi \frac{\partial F(t, x, \xi)}{\partial t}+f(\xi) \cdot \nabla_x F(t, x, \xi)\right) d \xi=0.
\end{align}

\begin{remark}
From a computational or numerical perspective, maximising entropy or minimizing energy makes the optimization convex and stable. It may eliminate high-frequency oscillations (common in high-resolution schemes) and ensure the convergence of the numerical scheme to the correct weak solution.
\end{remark}

Next we give a few physical examples and the corresponding concave entropies.
\medskip

\noindent{\bf Example 1}: 
Consider \emph{
the inviscid Burgers equation} with $f(u)=u^2/2$. Then one can take the entropy as 
\[
\eta(\xi) = -\xi^2.
\]
Maximizing entropy (which in this example corresponds to minimizing the energy) avoids oscillations and selects a Dirac measure corresponding to a physically relevant solution.
Furthermore, minimizing energy is equivalent to minimizing the variance of the solutions, thereby concentrating the measure. 
\medskip 

\noindent{\bf Example 2}:  
Consider \emph{the barotropic Euler system} with
\[
u = (\vr, \vm), \quad f(u)=\left( \vm, \ \frac{\vm \otimes \vm}{\vr} + p(\vr) \mathbb{I}_d \right), \quad p = \vr^{\gamma},
\]
where $\vr,\vm, p$ represent the density, momentum, and pressure of the fluid, respectively, and $\gamma$ is the adiabatic index. 
One can take the entropy as 
\begin{equation*}
    -\eta = \frac{|\vm|^2}{2\vr} + \frac{\vr^{\gamma}}{\gamma-1}.
\end{equation*}

Being inspired by the recent theoretical results of Feireisl, J\"ungel and Luk\'acov\'a \cite{Feireisl-Jungel-Lukacova}, our LP  problem \eqref{LP1}-\eqref{LP3} satisfies the selection criteria by minimizing the total energy
\begin{align*}
\mathcal{E} = \int_{\Omega} \int_{\mathbb{R}^n} (-\eta)(\xi) F(t, x, \xi) d \xi dx,
\end{align*} 
As shown in \cite{Feireisl-Jungel-Lukacova, Feireisl-Lukacova}, if there is a unique solution selected in  \eqref{LP1}-\eqref{LP3} then it is a unique weak solution of
the barotropic Euler system.

\medskip 

\noindent{\bf Example 3}: 
Consider \emph{the full Euler system} with the perfect gas law:
\[
u = (\vr, \vm, E), \ f(u)=\left( \vm, \ \frac{\vm \otimes \vm}{\vr} + p \mathbb{I}_d, \frac{\vm}{\vr} (E+p) \right), \ p = (\gamma-1)\vr e, \ E= \frac12 \frac{ |\vm|^2}{\vr}+ \vr e, 
\]
where $E,e$ represent the energy and specific internal energy of the fluid, respectively.
The entropy reads  
\begin{equation*}
    \eta = \vr \ln{(p/\vr^{\gamma})}.
\end{equation*}
Alternatively, one could also work with the governing equations for variables $(\vr, \vm, \eta)$ and minimize the energy. \\

In Appendix~\ref{app:simulations} we justify the mathematical validity of the LP formulations for three prototype nonlinear PDEs: the inviscid Burgers equation, the barotropic Euler equations, and the Allen-Cahn equations.

\begin{remark}
It is easy to generalize this linear programming method to a broader class of models, such as the Allen-Cahn equation and the Navier-Stokes equations. 
The efficiency of the method applied to the Allen-Cahn equation will be demonstrated through numerical simulations in Section~\ref{sec:allencahn}.
\end{remark}

\subsection{Discretization}\label{sec-det-dis}
We proceed to discretize the LP problem \eqref{LP1}-\eqref{LP3}. 
For the sake of simplicity, we discretize \eqref{LP1}-\eqref{LP3} applying a finite volume method together with the Lax-Friedrichs numerical flux.

Let $(t, x, \xi) \in [0, T] \times \Omega \times [\xi_{\min}, \xi_{\max}]^n$ be divided into uniform grid $\mesh_h$ with  $N_t N_x^dN_{\xi}^n$ cells. The discretization parameters in time, space, and phase space are denoted by $\TS, h_x$ and $h_{\xi}$, respectively.  We introduce the following notation 
\begin{align}\label{not}
\Ov{u}_l = \frac{1}{h_{\xi}^n}\int_{K_{\xi}^l} \xi d \xi, \quad  
\Ov{f}_l = \frac{1}{h_{\xi}^n}\int_{K_{\xi}^l} f(\xi) d \xi,  \quad 
\Ov{\eta}_l = \frac{1}{h_{\xi}^n}\int_{K_{\xi}^l} \eta(\xi) d \xi.
\end{align}
We seek 
\[
F_h = \frac{1}{h_{\xi}^n}  \sum_{j=1}^{N_t}\sum_{k}\sum_{l} F_{kl}^j 1_{K_t^j}1_{K_x^k}1_{K_{\xi}^l}
\]
for the linear programming problem \eqref{LP1}-\eqref{LP3} as follows:
\begin{align}
   &  \text{argmax}_F \sum_{k}\sum_{l} \Ov{\eta}_l F_{kl}^{j+1} 
\end{align}
subject to 
\begin{align}
& \sum_{l} F_{kl}^{j+1} =1, \qquad 
    F_{kl}^{j+1} \geq 0, \\ 
& \sum_{l} \left(\Ov{u}_l \frac{F_{kl}^{j+1}-F_{kl}^j}{\Delta t} + \sum_{i=1}^d \left[ (\Ov{f}_l)^{(i)} \frac{F_{k+e_i,l}^j-F_{k-e_i,l}^j}{2 h_x} - \varepsilon_{l} \Ov{u}_l \frac{F_{k+e_i,l}^j- 2 F_{k,l}^j +F_{k-e_i,l}^j}{h_x^2}  \right] \right)=0.  \label{eq:constraint1}
\end{align}
Here $k \in \{1, 2, \dots, N_x\}^d,l\in \{1, 2, \dots, N_{\xi}\}^n$ are vectors, $e_i$ is the unit vector in the positive $i$-th direction, the superscription $f^{(i)}$ represents the $i$-th component of $f$. 
Moreover, $\varepsilon$ is taken as
\begin{align}
\varepsilon_l = \frac12 h \rho(K_{\xi}^l), \ \mbox{ with } \ \rho: \mbox{ maximum absolute characteristic speed}.
\end{align}

\medskip

For each fixed $j \in \{1, 2, \dots, N_T\},$ we assemble a vector $F^j$ of size $N_x^d N_{\xi}^n$ with the entries $F_{kl}^j.$ To illustrate the ideas, we start with $d=1$:
\begin{enumerate}
\item 
\begin{align*}
    \sum_{k}\sum_{l} \Ov{\eta}_l F_{kl}^{j+1} = \mathbf{1}_{1\times N_x} \otimes \eta_{N_{\xi}^n} \cdot F^{j+1}, \quad \eta_{N_{\xi}^n} = (\Ov{\eta}_1, \dots, \Ov{\eta}_{N_{\xi}^n});
\end{align*}

\item 
\begin{align*}
    \sum_{l} F_{kl}^{j+1} = \mathbb{I}_{N_x} \otimes \mathbf{1}_{1\times N_{\xi}^n} \cdot F^{j+1};
\end{align*}

\item 
\begin{align*}
\mathbf{0}_{n N_x}= &\ \sum_{l} \left(\Ov{u}_l (F_{kl}^{j+1}-F_{kl}^j) +  \Ov{f}_l \frac{\TS}{2 h_x} (F_{k+1,l}^j-F_{k-1,l}^j)- \varepsilon_{l} \Ov{u}_l \frac{\TS}{h_x^2} (F_{k+1,l}^j- 2 F_{k,l}^j +F_{k-1,l}^j) \right) \\
= & \ \sum_{l} \Ov{u}_l F_{kl}^{j+1} + \sum_{l} ( a_l F_{k-1,l}^{j} + b_l F_{kl}^{j} + c_l F_{k+1,l}^{j}) \\
= & \ \mathbb{I}_{N_x} \otimes u_{N_{\xi}^n} \cdot F^{j+1} + \Big(\mathbb{J}_{N_x}(0))^T \otimes a_{N_{\xi}^n}  + \mathbb{I}_{N_x} \otimes b_{N_{\xi}^n} + \mathbb{J}_{N_x}(0) \otimes c_{N_{\xi}^n} + \mathbb{R}_{n N_x  \times N_x N_{\xi}^n } \Big) F^j
\end{align*}
with 
\begin{align*}
& a_l = - \Ov{f}_l \frac{\TS}{2 h_x} - \varepsilon_{l} \Ov{u}_l \frac{\TS}{h_x^2} , \quad b_l = -\Ov{u}_l + 2 \varepsilon_{l} \Ov{u}_l \frac{\TS}{h_x^2}, \quad c_l = \Ov{f}_l \frac{\TS}{2 h_x} - \varepsilon_{l} \Ov{u}_l \frac{\TS}{h_x^2}, \\
& X_{N_{\xi}^n} = (\Ov{X}_1, \dots, \Ov{X}_{N_{\xi}^n}), \quad X \in\{ u, a, b, c \}.
\end{align*}
\noindent Here, $\mathbb{J}_{N_x}(0)$ is the Jordan matrix and $\mathbb{R}_{n N_x \times N_x N_{\xi}^n }$ represents the boundary effects.
\end{enumerate}
Further, denoting
\begin{align*}
&  \widetilde{\Xi} = \mathbf{1}_{1\times N_x} \otimes \eta_{N_{\xi}^n}, \quad  
 \widetilde{A} = \mathbb{I}_{N_x} \otimes \mathbf{1}_{1\times N_{\xi}^n},\quad
 \widetilde{B} = \mathbb{I}_{N_x} \otimes u_{N_{\xi}^n}, \\
& \widetilde{D} = \mathbb{J}_{N_x}(0)^T \otimes a_{N_{\xi}^n}  + \mathbb{I}_{N_x} \otimes b_{N_{\xi}^n} + \mathbb{J}_{N_x}(0) \otimes c_{N_{\xi}^n} + \mathbb{R}_{n N_x \times N_x N_{\xi}^n }, 
\end{align*}
we rephrase the linear programming problem \eqref{LP1}-\eqref{LP3} into the following matrix formulation:
\begin{align}
& \text{argmax}_{\mathbf{F}^{j+1}} \widetilde{\Xi} \mathbf{F}^{j+1}, \nonumber \\
& \mathbf{F}^{j+1} \geq \mathbf{0}_{N_xN_{\xi}^n \times 1}, \quad \widetilde{M}\mathbf{F}^{j+1}= \mathbf{c}_j, \quad 
\widetilde{M}=\begin{pmatrix}
            \widetilde{A} \\
            \widetilde{B}
\end{pmatrix}, \quad 
\mathbf{c}_j = \begin{pmatrix}
            \mathbf{1}_{N_x \times 1}\\
            -\widetilde{D} \mathbf{F}^j := \mathbf{d}_j
\end{pmatrix}, 
\end{align}
or
\begin{align} \label{eq:mainLP}
& \text{argmax}_{\mathbf{F}} \, {\Xi}\, \mathbf{F}, \nonumber \\
& \mathbf{F} \geq \mathbf{0}_{N_tN_xN_{\xi}^n \times 1}, \quad {M}\mathbf{F}= (\mathbf{1}_{N_x \times 1}; \mathbf{0}_{n N_x \times 1}), 
\end{align}
with
\begin{align*}
& {M}=\begin{pmatrix}
            {A} \\
            {B}
\end{pmatrix}, \quad 
{\Xi} = \mathbf{1}_{1\times N_t} \otimes \widetilde{\Xi}, \quad 
{A} = \mathbb{I}_{N_t} \otimes \widetilde{A}, \quad 
{B} = \mathbb{J}_{N_t}(0)^T \otimes \widetilde{D} +  \mathbb{I}_{N_t} \otimes \widetilde{B}.
\end{align*}

It is easy to see that the matrices $\widetilde{\Xi}, \widetilde{A}, \widetilde{B}, \widetilde{D}$ are independent of $j$ and obtained once $[\xi_{\min}, \xi_{\max}]$ and $N_{\xi}$ are given. Moreover, $\widetilde{A}, \widetilde{B}, \widetilde{D}$ have a sparsity structure of $N_{\xi}^n, \max(n, N_{\xi}^n)$ and $3\max(n, N_{\xi}^n)$, respectively. 
Consequently, $\widetilde{M}$ is a $r \times s$ matrix with $r=N_x (1 + n)$ rows and $s=N_x N_{\xi}^n$ columns. The total number of non-zero elements of $\widetilde{M}$ is $ N_x N_{\xi}^n (n+1)$.

\medskip

Correspondingly, $\Xi, A, B$ are independent of $j$ and obtained once $[\xi_{\min}, \xi_{\max}]$ and $N_{\xi}$ are given. Moreover, $A, B $ have a sparsity structure of $N_{\xi}^n$ and $4\max(n, N_{\xi}^n)$, respectively. 
Consequently, $M$ is a $r \times s$ matrix with $r=N_t N_x (1 + n)$ rows and $s=N_t N_x N_{\xi}^n$ columns. The total number of non-zero elements of $M$ is $ N_t N_x N_{\xi}^n (4n+1)$.

\bigskip
Now we generalize $d=1$ to $d>1$ with
\begin{align*}
&  \widetilde{\Xi} = \mathbf{1}_{1\times N_x^d} \otimes \eta_{N_{\xi}^n}, \quad  
 \widetilde{A} = \mathbb{I}_{N_x^d} \otimes \mathbf{1}_{1\times N_{\xi}^n},\quad
 \widetilde{B} = \mathbb{I}_{N_x^d} \otimes u_{N_{\xi}^n}, \\
& \widetilde{D} = \sum_{i=1}^d \left( \mathbb{K}_{N_x^d}^i(0)^T \otimes a_{N_{\xi}^n}  + \mathbb{I}_{N_x^d} \otimes b_{N_{\xi}^n} + \mathbb{K}_{N_x^d}^i(0) \otimes c_{N_{\xi}^n}  + \mathbb{R}^i_{N_x^d n \times N_x^d N_{\xi}^n} \right), \\
& \mathbb{K}^i_{N_x^d}(0) = \bigotimes_{k < i} \mathbb{I}_{N_x} \otimes \mathbb{J}_{N_x}(0)  \otimes \bigotimes_{k>i} \mathbb{I}_{N_x}. 
\end{align*}
Therefore, we have that
$M$ is a $r \times s$ matrix with $r=N_t N_x^d (1 + n)$ rows and $s=N_t N_x^d N_{\xi}^n$ columns. The total number of non-zero elements of $M$ is $ N_t N_x^d N_{\xi}^n \Big[ 2(d+1)n+1 \Big]$.

\subsection{Numerical justification}
We conclude this section with a numerical justification of the LP problem \eqref{LP1}-\eqref{LP3} and its discrete counterpart \eqref{eq:mainLP}. 

To this end, we have performed one-dimensional simulations on classical computers for three prototype nonlinear PDEs: the inviscid Burgers equation, the barotropic Euler equations, and the Allen-Cahn equation.
The first two equations are nonlinear hyperbolic conservation laws, while the Allen-Cahn equation is a reaction-diffusion equation.
Thanks to the well-known results on the existence and uniqueness of weak entropy solutions for the above three (one-dimensional)  nonlinear PDEs, it is expected that our mathematical model - Young-measure based LP problem - will recover the unique weak entropy solution. 

In the numerical simulations, we present the numerical errors, comparisons with an exact (or a reference) solution, and defects. The latter represents oscillations of a truly measure-valued solution. As we illustrate in the Appendix, the defects are of the order of $10^{-4}$ (due to discretization errors in phase space), indicating that the measure obtained by \eqref{eq:mainLP} is a Dirac measure, i.e., the measure-valued solution is the weak entropy solution. Moreover, our numerical solutions are accurate approximations of the exact solutions.  
Overall, these observations validate our mathematical model. 
For completeness, we present the results of numerical simulations in  Appendix~\ref{app:simulations}.

\section{Quantum algorithms to compute the Young measures for deterministic nonlinear PDEs} \label{sec:youngnonlinearpde1}

Thanks to the linearity of the LP formulation using the Young measure, it is natural to see whether one can develop efficient quantum algorithms for the original nonlinear PDEs based on the LP formulation in order to overcome the curse of dimensionality. Two questions naturally arise. First, is the mathematical formulation valid for general nonlinear PDEs?  For three prototype examples -- the inviscid Burgers equations, the Barotropic Euler equations, and the Allen-Cahn equation -- we give numerical results in one space dimension in Appendix~\ref{app:simulations}.

While we have not yet demonstrated the validity of the LP approach for the full Euler and Navier-Stokes equations, we also include these examples in our comparison of computational costs to illustrate the complexities of potential future applications.  \\

The second question is: Does this formulation yield quantum algorithms that offer advantages compared to their classical counterparts?    It is the aim of this section to examine the cost of solving~\eqref{eq:mainLP} with quantum algorithms and their comparison to classical LP methods.  

\subsection{Computational complexities of QLP algorithms}
\label{Sec:cost}
 The majority of quantum algorithms for LP outputs the classical vector $\mathbf{F}$. While they are mostly built on quantum versions of the interior-point method (IPM), there are also other approaches, such as quantum zero-sum games and the quantum central path (QCP) method, which have no direct classical analog. For an LP problem with $r$ constraints and $s$ variables, see Table~\ref{tab:1} for a summary of the cost of classical and some different classes of quantum algorithms. To apply this to our problem of computing Young measures, see Table~\ref{tab:2} for the relationship of $r$, $s$ and other properties of the LP problem in terms of the physical dimension $d$, the size $n$ of the original nonlinear PDE system, and discretisation sizes $N_t$, $N_x$ and $N_{\xi}$.  \\

\begin{table}[h] 
\centering
\caption{Summary of cost of classical and some different classes of quantum algorithms for LP problems, where $\tilde{O}$ hides logarithmic factors in the number of constraints $r$, number of variables $s$, and precision $\epsilon$. The quantum cost is expressed in terms of query complexity, but the types of oracles used vary across algorithms. Here $\mathbf{F}$ denotes the classical output of the LP solution $\mathbf{F}$ and $R_1=\|\mathbf{F}\|_1$. The density matrix containing $\mathbf{F}$ along its diagonal is denoted $\rho_F^*$. The number of non-zero elements in the constraint matrix $M$ in the primal LP problem is denoted $\text{nnz}(M)$, and $s_M$ is the maximum of the row and column sparsity of $M$. $\kappa_{Newt}$ is the condition number of the matrix to be inverted at each Newton step of the IPM.
Here and hereafter, Stochastic IPM, Quantum IPM, Quantum zero-sum-game, Quantum central path, Quantum SDP are denoted by SIPM, QIPM, QZSG, QCP, QSDP, respectively.} \label{tab:1}
\begin{tabular}{|l|l|l|l|ll} 
\cline{1-4}
Algorithm  &  Time/query complexity & Output & Comments  &  \\ \cline{1-4}
IPM \cite{lee2015efficient} & $\tilde{O}(\sqrt{s}(\text{nnz}(M)+s^2))$ &  $\mathbf{F}$  & $\log(1/\epsilon)$ dependence  &  \\ \cline{1-4}
SIPM \cite{cohen2021solving} & $\tilde{O}(s^{\omega})$, \quad $\omega \sim 2.38$ &   $\mathbf{F}$  &  $\log(1/\epsilon)$ dependence  &  \\ \cline{1-4}
QIPM \cite{kerenidis2020quantum} & $\tilde{O}(\kappa^3_{Newt} s^{2}/\delta^2)$  &  $\mathbf{F}$ & Error $\delta$ 
for tomography &  \\ \cline{1-4}
QZSG \cite{van2019quantum} & $\tilde{O}(\sqrt{s_M} (R_1/\epsilon)^{3.5})$ & $\mathbf{F}$ &  $\inf \|\lambda\|_1\sim \tilde{O}(1)$ &  \\ \cline{1-4}
QCP \cite{augustino2023quantum}& $\tilde{O}(\sqrt{r+s}R_1/\epsilon)$ &  $\mathbf{F}$ & Gate complexity: query $\times \text{nnz}(M)$ &  \\ \cline{1-4}
QSDP \cite{liu2025quantum} &  $\tilde{O}(rR_1^2s /\epsilon^2)$  & $\rho^*_F$ &  $\rho_F$ prep cost not incl. &  \\ \cline{1-4}
\end{tabular}
\end{table}

\begin{table}[h] 
\centering
\caption{Sizes of parameters related to LP problem in~\eqref{eq:mainLP}, where equalities are stated in terms of $\tilde{O}$ for simplicity, which denotes that all terms logarithmic or smaller in $r$, $s$ are suppressed, so no multiplicative dependence on $d$, $n$ is included. Here, we express these parameters in terms of the discretization sizes and dimensions of the original formulation of the problem, as well as in terms of a modification based on a particle method. Here $\mathbf{F}$ denotes the classical output of the LP solution $\mathbf{F}$ and $R_1=\|\mathbf{F}\|_1$.  The number of non-zero elements in the constraint matrix $M$ in the primal LP problem is denoted $\text{nnz}(M)$, and $s_M$ is the maximum of the row and column sparsity of $M$. } \label{tab:2} 
\begin{tabular}{|l|l|l|ll} 
\cline{1-3}
Parameter  &  LP-FxM $\tilde{O} (\cdot)$& LP-Particle method $\tilde{O}(\cdot)$ &  \\ \cline{1-3}
$r$ & $N_tN_x^d$ & $N_t N_x$ & \\ \cline{1-3}
$s$ & $rN_{\xi}^n=N_t N_x^d N_{\xi}^n$ & $rN_{\xi}=N_t N_x N_{\xi}$ &   \\ \cline{1-3}
$R_1$ & $r=N_t N_x^d$ & $r=N_t N_x$ &  \\ \cline{1-3}
$\text{nnz}(M)$  & $s=N_t N_x^dN_{\xi}^n$ & $s=N_t N_x N_{\xi}$ &  \\ \cline{1-3}
$s_M$ & $N_{\xi}^n$ & $N_{\xi}$ &\\ \cline{1-3}
\end{tabular}
\end{table}

 The quantum interior-point methods (QIPMs) for
LP largely retains the classical primal-dual IPM architecture, which relies on $\tilde{O}(\sqrt{s})$ (where there is a logarithmic dependence on $1/\epsilon$) iterations of the primal-dual Newton's steps. In the quantum algorithm, this outer path-following loop and its $\tilde{O}(\sqrt{s})$ iteration complexity remain essentially unchanged, while the main quantum ingredient is to replace the Newton's step at each iteration by a quantum linear systems algorithm (QLSA), whose cost depends on the condition number $\kappa_{Newt}$ of the matrix that needs to be inverted in each iteration (see Appendix~\ref{app:1}). This is followed by quantum state tomography to recover a classical approximation of the Newton direction before the next iterate is formed \cite{kerenidis2020quantum, casares2020quantum}. In this sense, current QIPMs are hybrid quantum-classical rather than fully quantum: the potential advantage comes from accelerating the inner linear algebra, but the need to tomographically read out the search direction -- together with dependence on conditioning and strong data-access assumptions such as block-encodings or QRAM -- often erodes the asymptotic gain in end-to-end settings. In \cite{kerenidis2020quantum} the worst-case end-to-end cost (including tomography) is claimed to be $\tilde{O}(\kappa_{Newt}^3 s^{1.5} \mu/\delta^2)$, where $\mu \leq \sqrt{2s}$ and $\delta$ is the precision in measurement in each tomography step. The estimation of the condition number $\kappa_{Newt}$ is required for each case and would be a direction for future work in this direction to diminish the dependence on condition number. Subsequent work has focused less on changing the outer IPM logic and more on making the Newton-step more quantum-compatible, for example, by using inexact-feasible or inexact-infeasible formulations, iterative refinement, and improved error management so that fewer or less accurate tomography calls are needed while preserving classical convergence guarantees \cite{qipmreview}. There have been developments on QIPM beyond QLSA and tomography, for example, in \cite{apers2023quantum}, where the cost does not depend on the condition number $\kappa_{Newt}$. However, here the quantum advantage only applies to 'tall' LPs ($r \gg s$), which is not the regime we have in our LP problem. \\

 For LP, two notably different quantum routes are the {\it quantum central path (QCP)} method and quantum zero-sum-games. QCP encodes the entire self-dual central path into a time-dependent Hamiltonian, followed by a single continuous evolution step, so the solution is obtained in a single final measurement rather than by repeated Newton steps. It is particularly notable for avoiding the iterative QLSA + tomography loop that burdens the previous hybrid QIPMs. For LPs, it has query complexity $\tilde{O}(\sqrt{r+s} R_1/\epsilon)$ and gate complexity $\tilde{O}(\sqrt{r+s} R_1 \text{nnz}(M)/\epsilon)$  without QRAM, while recovering an exactly feasible primal-dual pair up to the stated optimization error. Here $R_1=\|\mathbf{F}\|_1$  and $\text{nnz}(M)$  is the number of non-zero elements of matrix $M$. For our problem applied to PDEs, the normalisation of $\mathbf{F}$ in $l_1$-norm is $R_1=\tilde{O}(N_t N_x^d)$, which comes from the probability interpretation of $\mathbf{F}$ with respect to $\xi$, while $\text{nnz}(M)=\tilde{O}(s)$. This means that even without considering $\kappa_{Newt}$ in QIPM, the QCP method has better scaling in both $r$ and $\cmag n$, as seen in Table~\ref{tab:1} and Table~\ref{tab:3}. In fact, compared with classical IPM methods, the query complexity of QCP has an advantage in $r$ by up to a factor of $r$, where $r$ scales exponentially with the physical dimension $d$. It also has \textit{polynomial advantage} in $N_\xi$, by a factor of $N_{\xi}^{2n}$. The dependence of the cost on precision $\epsilon$, however, is greater than the $\log(1/\epsilon)$ in the classical algorithm. If we instead compare classical cost with the gate complexity in QCP, there is only favorable scaling with respect to $N_{\xi}$ but not with respect to $r$. For example, if an $k$-th order numerical quadrature is used to discretize the integral in $\xi$, then the total quantum advantage is $N_{\xi}^{n}\epsilon=O(N_{\xi}^{n-k})$, so this advantage can be large when $n \gg k$. \\

 By contrast, the quantum zero-sum-game methods (QZSG) first reduce the LP to a matrix zero-sum-game and then use quantum Gibbs sampling and multiplicative-weights-style updates. This gives sublinear dependence on the problem size. In particular, in the case when  $M$ is sparse, with sparsity $s_M$, the cost is $\tilde{O}\left(\sqrt{s_m} ((R_1(\inf \|\lambda\|_1+1)/\epsilon)^{3.5}\right)$, where $\lambda$ is the $r-$ dimensional Lagrange multiplier vector in the dual LP problem. In the PDE problem, this $\inf \|\lambda\|_1$ parameter can be considered an additional stability parameter of the discretised LP problem, where in the stable and well-behaved LP and PDE case one would not expect to be exponential in $d$ or $D$ and could even be $O(1)$, so here we use $\inf \|\lambda\|_1=\tilde{O}(1)$. However, its exact behavior for each PDE should be further investigated. Unlike in QCP, this complexity comes at the price of stronger QRAM-style input assumptions, and the returned primal-dual pair is only approximately feasible. One can see from Table~\ref{tab:3} that its complexity is not favorable in terms of $r$ compared to classical methods, but, like QCP, has {\it polynomial advantage} in terms of $N_{\xi}$, namely by a factor of $N_{\xi}^{2n}$. However, the dependence on both $r$ and $1/\epsilon$ is higher for QZSG.\\

While the quantum algorithms above directly return the classical output $\mathbf{F}$, in principle it is also possible to prepare quantum states that embed the solution $\mathbf{F}$. One notable example is quantum algorithms for SDPs \cite{brandao2017quantum, liu2025quantum}, where LPs are a special instance of Semi-definite Programming (SDP), where the SDP problem is to find $\min_{\rho_F} \text{Tr}(C\rho_F)$ for matrix $\rho_F$ subject to the constraints $\text{Tr}(Q_i \rho_F)=q_i$, $i=1, \cdots, r$. These algorithms are built on preparing quantum thermal states $\rho_F$, and there are $r$ classical parameters (considered as chemical potentials) to be updated classically \cite{liu2025quantum}. The typical way to embed LP into SDP is by embedding $\mathbf{F}$ into the diagonal of the matrix $\rho_F$. However, if $\rho_F$ appears to be diagonal, this could be generated by a diagonal Hamiltonian, thus one would not be exploiting quantum coherence. Then it is unclear what quantum advantage this would provide, if any, and it remains to be further investigated. If the thermal state preparation is considered as an oracle, then there are $r$ parameter updates with $\tilde{O}(R_1^2 s/\epsilon^2)$ iterations, so a total query complexity is of $\tilde{O}(rR_1^2s/\epsilon^2)=\tilde{O}(r^4N_{\xi}^n/\epsilon^2)$. Compared to the classical methods in Table~\ref{tab:3}, we see that this could be potentially favorable only in terms of $N_{\xi}$-dependence, but at the cost of greater $ r$-dependence. Thus, in Table~\ref{tab:3}, we list only algorithms with $\mathbf{F}$ as the output; we do not list QSDP there, and we leave the details of QSDP for future study. \\

\begin{table}[h] 
\centering
\caption{Comparison of costs to output classical $\mathbf{F}$ for different algorithms in terms of $r=N_tN_x^d$, $n$ and discretisation size $N_{\xi}$.} \label{tab:3}
\begin{tabular}{|l|l|l|ll} 
\cline{1-3}
Algorithm  &  Time/query complexity  & Comments  &     \\ \cline{1-3}
IPM \cite{lee2015efficient} & $\tilde{O}(r^{2.5}N_{\xi}^{5n/2})$ & $\log(1/\epsilon)$ dependence  &     \\ \cline{1-3}
SIPM \cite{cohen2021solving} & $\tilde{O}(r^{2.38}N_{\xi}^{2.38n})$ &  $\log(1/\epsilon)$ dependence  &     \\ \cline{1-3}
QIPM\cite{kerenidis2020quantum} & $ \tilde{O}(\kappa^3_{Newt} r^{2} N_{\xi}^{2n}/\delta^2)$  & $\kappa_{Newt}$ bound to be determined &   \\ \cline{1-3}
QZSG\cite{van2019quantum} & $\tilde{O}((r/\epsilon)^{3.5}N_{\xi}^{n/2})$ & Assuming $\inf \|\lambda\|_1\sim \tilde{O}(1)$ &     \\ \cline{1-3}
QCP \cite{augustino2023quantum}& $\tilde{O}(r^{1.5}N_{\xi}^{n/2}/\epsilon)$ & Gate complexity $\tilde{O}(r^{2.5}N_{\xi}^{3n/2}/\epsilon)$ &    \\ \cline{1-3}
\end{tabular}
\end{table}

From Table~\ref{tab:3} one can see that quantum algorithms are often more favorable in terms of the cost-dependence on $N_{\xi}$, where the polynomial advantage in $N_{\xi}$ depends on $n$.  This suggests that one can expect greater quantum advantage for problems where $n \gg 1$.  For fluid equations (as can be seen from Table~\ref{tab:4}), one can have $n=d+2$, but here one is restricted to $d_{\max}=3$, so the quantum advantage for fluid dynamics is not that large. Examples, where $n$ can be extremely large, include cases where one studies chemically reacting flows of many chemical species, with $n$ corresponding to the number of chemical species. \\

Another extension is to consider a particle-methods reformulation of the LP problem, where instead of using $s=N_t N_x^d N_{\xi}^n$ grid points for $\mathbf{F}$, we use $r$ particles that traverse the spatial dimensions in $x$-space and $s/r$ particles that traverse the dimensions in $\xi$-space, so $r$ and $s$ no longer depend on the number of physical dimensions. See Table~\ref{tab:2} for how $r$, $s$, and other parameters depend on the discretization sizes, and these factors no longer have exponential dependence on $d$ or $n$. Simple estimates using Table~\ref{tab:2} and Table ~\ref{tab:1} show that the expected quantum advantages are in fact much less when using a particle-methods approach, as those advantages would not grow as $d$ and $n$ grow. \\

Suppose we go beyond the first-order methods, and use, e.g., higher-order FxMs (finite element, finite difference, and finite volume methods) ($1/\epsilon \sim N^k$ for $k^{\text{th}}$-order methods) and spectral methods ($1/\epsilon \sim e^N$). If we let discretisation size $N \sim N_x \sim N_{\xi} \sim N_t$ for simplicity, we can represent the classical cost by $O(N^c)$ (neglecting the logarithmic factor $\log(1/\epsilon)$) and let $O(N^{q}/\epsilon)$ represent the quantum cost, so quantum advantage can be expressed in terms of $O(N^{c-q} \epsilon)$. If we re-express in terms of discretization size alone, we see that $\epsilon \sim N^{-k}$ for FxM, thus $O(N^{c-q}/\epsilon)=O(N^{c-q-k})$ becomes much smaller as $k$ grows, and the disadvantage of the quantum scaling is worse for spectral methods. Thus, we see that the current quantum advantages in $N$ are expected to be smaller and even vanish as $k$ grows. This is due to the more favorable scaling of classical methods with precision, $\log(1/\epsilon)$, compared to quantum algorithms, which, even in the best-case scenario, have scaling of $1/\epsilon$. The latter is true if we want the quantum algorithm to recover a classical $\mathbf{F}$ output, since the cost in sampling from quantum states cannot exceed $1/\epsilon$. \\

Since this $1/\epsilon$ or $1/\epsilon^2$ scaling with precision is inevitable for quantum algorithms that are expected to output classical results $\mathbf{F}$, one can speculate on possible advantages that can be obtained in developing quantum algorithms that output instead a quantum state embedding $\mathbf{F}$. This would have two potential advantages: (i) to decrease the dependence on $1/\epsilon$ and (ii) to have potential advantages in the computation of moments of the averaged solutions $\bar{u}^m=\int \xi^m F(t, x, \xi) d\xi^n$. The latter has been done using the {\it Lasserre hierarchy} \cite{cardoen2024},  which can be considered as an inner product. For example, in realistic scenarios, we would be interested in the first and second moments of $F$, such as density, velocity, and energy for the fluid equations. If one prepares a quantum state $|F\rangle \propto \sum_{i,j,k} F(t_i, x_j, \xi_k)|i\rangle |j\rangle |k\rangle$, then $\bar{u}^m$ could be estimated by a state-overlap algorithms. like quantum swap-tests, which are up to exponentially more efficient than computing classical inner products, since one is essentially taking an inner product with a state of size $\log(N_t N_x^d N_{\xi}^n)$. One must be careful of the impact of normalization constants, so a more careful study is required. If one instead prepares a quantum state $|\sqrt{F}\rangle \propto \sum_{i,j,k} \sqrt{F(t_i, x_j, \xi_k)}|i\rangle |j\rangle |k\rangle$, then $\bar{u}^m$ can be recovered instead by quantum sampling, and here one can have up to {\it polynomial} advantage. To solve nonlinear PDEs, recovering a final classical $\mathbf{F}$ is not always strictly necessary unless one wants Young measures. Indeed, we are typically interested to recover the related observables/moments (expected values with respect to the Young measure or variances/defects). In such a situation, the quantum algorithms for outputting quantum states that contain $F$ could be more appropriate and would be an important direction for future research.\\

Overall, the current quantum LP algorithms, such as QCP, appear most promising for computing Young measures when $n$ is large; see Table~\ref{tab:3}, which shows quantum advantage for first-order methods of order $\tilde{O}(N_{\xi}^{n-1})$. However, while current quantum algorithms appear promising for computing the Young measures associated with nonlinear PDEs when $n$ is large, the conclusion is different if we instead compare the complexity of the quantum LP algorithm with that of {\it direct} classical solvers for the original nonlinear PDEs. For direct classical solvers of a system of $n$ nonlinear PDEs in physical dimension $d$, the cost is of order $O(nN_tN_x^d)=\tilde{O}(r)$. Comparing this to the costs in Table~\ref{tab:3} (without even including the extra cost in recovering the measure-valued solution from $\mathbf{F}$), we see that none of the quantum LP algorithms can improve upon the direct classical solvers for the observables, the expected value $\bar{u}$ of the measure-valued solution. Thus, further investigation in this direction -- either finding improved quantum LP algorithms; constructing new quantum LP algorithms by constructing quantum states like $|F\rangle$ or $|\sqrt{F}\rangle$ embedding the Young measure (to get quantum advantage in $1/\epsilon$ when recovering moments $\bar{u}^m$); and/or utilizing the special structure of the underlying mathematical model
to reduce the cost of the current quantum LP algorithms 
-- is needed.

\subsection{Computing the Young measures for three prototype examples}
 In Table~\ref{tab:4}, we have summarised the costs to recover the Young measures for the inviscid Burgers, Allen-Cahn, Euler, and the Navier-Stokes equations in terms of the discretization size and the physical dimension $d$. In Table~\ref{tab:5}, we restrict $d=3$, from which we see that the quantum central path (QCP) method provides the most promising direction for quantum advantage to output $\mathbf{F}$. For simplicity, assuming first order methods, $N_t \sim N_x \sim N_{\xi} \sim 1/\epsilon=N$, then comparing the query complexity of QCP with time complexity of stochastic IPM, there is a factor $\sim N^{4.4}$ quantum advantage for inviscid Burgers and Allen-Cahn equations, a factor $\sim N^{10.04}$ quantum advantage for barotropic Euler's equations and a factor $\sim N^{11.92}$ quantum advantage for the full Euler's and Navier-Stokes equations, which is a more than quadratic advantage. If we compare stochastic IPM with the gate complexity of QCP on the other hand, there is no quantum advantage for inviscid Burgers (see \cite{hjviscosityarxiv} for a different quantum algorithm for inviscid Burgers' that is globally valid in time) or Allen-Cahn equations, in fact the quantum gate complexity requires an extra factor $\sim N^{0.6}$ compared to the classical stochastic IPM. For barotropic Euler's equations, there is a more modest quantum advantage of $\sim N^{2.04}$, and for the compressible Euler and the Navier-Stokes equations, there is a potential quantum advantage by a factor $\sim N^{2.92}$.\\

While these advantages appear quite modest for recovering the Young measure, they are nevertheless non-trivial and are the first quantum algorithms to show quantum advantage for problems related to nonlinear fluid dynamical equations that remain globally valid in time, thus capable of capturing truly nonlinear and global properties. Since this is only a first exploration, these numbers are only a rough indication at this point, and further studies are required to study these algorithms and their potential improvements in more depth. \\

However, it is important to emphasize that none of these quantum methods are currently  advantageous compared to the cost of classical algorithms that directly solve the original PDEs (instead of the Young measures), where the complexity of the classical algorithms is $O(dN_t N_x^d)$. In this paper, we focus only on the problem of recovering the Young measure - so the comparison is with classical LP methods - which is an important problem in its own right.

\begin{table}[h] 
\centering
\caption{Time/query complexity for different nonlinear PDEs (to compute the Young measures), in terms of discretization sizes $N_t$, $N_x$, and $N_{\xi}$ and physical dimension $d$. For the inviscid Burgers and Allen-Cahn equations, $n=1$. For the barotropic Euler equations, $n=d+1$. For the full compressible Euler equations and the Navier-Stokes-Fourier (NSF) equations, $n=d+2$.}
\label{tab:4}
\begin{tabular}{|l|l|l|l|ll} 
\cline{1-4}
Algorithm  & Burgers; Allen-Cahn& Barotropic Euler  & Full Euler; NSF \\ \cline{1-4}
IPM  & $N_t^{2.5} N_x^{2.5d}N_{\xi}^{2.5}$ &  $N_t^{2.5} N_x^{2.5d}N_{\xi}^{2.5(d+1)}$  & $N_t^{2.5} N_x^{2.5d}N_{\xi}^{2.5(d+2)}$ &  \\ \cline{1-4}
SIPM  & $N_t^{2.38} N_x^{2.38d}N_{\xi}^{2.38}$&  $N_t^{2.38} N_x^{2.38d}N_{\xi}^{2.38(d+1)}$   &  $N_t^{2.38} N_x^{2.38d}N_{\xi}^{2.38(d+2)}$  &  \\ \cline{1-4}
QIPM & $\kappa^3_{Newt} N_t^2 N_x^{2d} N_{\xi}^{2}/\delta^2$  & $\kappa^3_{Newt} N_t^2 N_x^{2d} N_{\xi}^{2(d+1)}/\delta^2$  & $\kappa^3_{Newt} N_t^2 N_x^{2d} N_{\xi}^{2(d+2)}/\delta^2$ &  \\ \cline{1-4}
QZSG & $N_t^{3.5}N_x^{3.5d}N_{\xi}^{0.5}/\epsilon^{3.5}$ & $N_t^{3.5}N_x^{3.5d}N_{\xi}^{0.5(d+1)}/\epsilon^{3.5}$ &  $N_t^{3.5}N_x^{3.5d}N_{\xi}^{0.5(d+2)}/\epsilon^{3.5}$ &  \\ \cline{1-4}
QCP& $N_t^{1.5}N_x^{1.5d}N_{\xi}^{0.5}/\epsilon$ &  $N_t^{1.5}N_x^{1.5d}N_{\xi}^{0.5(d+1)}/\epsilon$ & $N_t^{1.5}N_x^{1.5d}N_{\xi}^{0.5(d+2)}/\epsilon$ &  \\ \cline{1-4}

\end{tabular}
\end{table}

\begin{table}[h] 
\centering
\caption{Time/query complexity for different nonlinear PDEs (for the Young measures), in terms of discretization sizes $N_t$, $N_x$, and $N_{\xi}$ when dimension $d=3$.}
\label{tab:5}
\begin{tabular}{|l|l|l|l|ll} 
\cline{1-4}
Algorithm  & Burgers; Allen-Cahn& Barotropic Euler  & Full Euler; NSF \\ \cline{1-4}
IPM  & $N_t^{2.5} N_x^{7.5}N_{\xi}^{2.5}$ &  $N_t^{2.5} N_x^{7.5}N_{\xi}^{10}$  & $N_t^{2.5} N_x^{7.5}N_{\xi}^{12.5}$ &  \\ \cline{1-4}
SIPM  & $N_t^{2.38} N_x^{7.14}N_{\xi}^{2.38}$&  $N_t^{2.38} N_x^{7.14}N_{\xi}^{9.52}$   &  $N_t^{2.38} N_x^{7.14}N_{\xi}^{11.9}$  &  \\ \cline{1-4}
QZSG & $N_t^{3.5}N_x^{10.5}N_{\xi}^{0.5}/\epsilon^{3.5}$ & $N_t^{3.5}N_x^{10.5}N_{\xi}^{2}/\epsilon^{3.5}$ &  $N_t^{3.5}N_x^{10.5}N_{\xi}^{2.5}/\epsilon^{3.5}$ &  \\ \cline{1-4}
QCP (query)& $N_t^{1.5}N_x^{4.5}N_{\xi}^{0.5}/\epsilon$ &  $N_t^{1.5}N_x^{4.5}N_{\xi}^{2}/\epsilon$ & $N_t^{1.5}N_x^{4.5}N_{\xi}^{2.5}/\epsilon$ &  \\ \cline{1-4}
QCP (gate)& $N_t^{2.5}N_x^{7.5}N_{\xi}^{1.5}/\epsilon$ &  $N_t^{2.5}N_x^{7.5}N_{\xi}^{6}/\epsilon$ & $N_t^{2.5}N_x^{7.5}N_{\xi}^{7.5}/\epsilon$ &  \\ \cline{1-4}
\end{tabular}
\end{table}

\section{Quantum algorithms to compute the Young measures for random  nonlinear PDEs} \label{sec:uncertain}

In this section, we consider a random system of  nonlinear hyperbolic conservation laws
\begin{eqnarray}
\label{RPDE1}
&& \partial_t u(t,x,\omega) + \Div f(u(t,x,\omega))  = 0,   \qquad   (t,x,\omega) \in (0,T)\times \Omega \times D,\nonumber \\ 
&& u(0,x,\omega)=u_0(x,\omega) \in \Bbb{R}^n.
\end{eqnarray}
For simplicity, we use periodic boundary conditions and assume that the randomness enters only through the random initial data $u_0$, which is a Borel measurable mapping from $D \subset \mathbb{R}^{m}$ to the phase space in $\mathbb{ R}^n$ with probability density $p(\omega)$.
In addition to \eqref{RPDE1} we look for a (measure-valued) solution satisfying the entropy inequality, i.e.
\begin{align}
\label{RPDE2}
\partial_t \eta(u(t,x,\omega)) + \Div q(u(t,x,\omega))  \geq  0.
\end{align}

\subsection{Linear programming problem}\label{sec:LP-Random}

We start by recalling the definition of the measure-valued solution of \eqref{RPDE1}-\eqref{RPDE2} in \cite{chu2025solving}, which is a natural extension of the definition for the deterministic problem. 
\begin{definition}\label{def-RPDE}
    The Young measure $\mathcal{V}_{t,x,\omega} \in \mathcal{P}(\mathbb{R}^n)$ is a measure-valued solution of \eqref{RPDE1}-\eqref{RPDE2} with
    \begin{align}\label{Rinitial-ym-0}
\mathcal{V}_{0,x,\omega}(\cdot) = \delta(\cdot - u_0(x,\omega) ) \mbox{ for  a.a. }  x \in \Omega, \ \omega \in D,
\end{align}
if 
\begin{align}\label{RPDE-YM}
\int_{0}^T \int_{\Omega} \int_D \left( \partial_t \psi(t,x)  \int_{\mathbb{R}^n} \u \;  \mathrm{d}\mathcal{V}_{t,x,\omega}(\u)   + \Grad  \psi(t,x) : \int_{\mathbb{R}^n} f(\u) \; \mathrm{d}\mathcal{V}_{t,x,\omega}(\u) \right) \phi(\omega) p(\omega)  d \omega d x d t  = 0,
\end{align}
for all test functions $\psi \in C^1_c((0,T) \times \Omega; \Bbb R^n),$ $\phi \in C(D) $; and 
\begin{align} \label{RPDE-YM-entropy}
\int_{0}^T \int_{\Omega} \int_D \left( \partial_t \tilde{\psi}(t,x)  \int_{\mathbb{R}^n} \eta(\u) \;  \mathrm{d}\mathcal{V}_{t,x,\omega}(\u)   + \Grad  \tilde{\psi}(t,x) \cdot \int_{\mathbb{R}^n} q(\u) \; \mathrm{d}\mathcal{V}_{t,x,\omega}(\u)  \right)\tilde{\phi}(\omega) p(\omega) d \omega d x d t  \leq 0,
\end{align}
for all test functions $\tilde{\psi} \in C^1_c((0,T) \times \Omega),$ $\tilde{\phi} \in C(D) $, $\tilde{\psi} \geq 0, \tilde{\phi} \geq 0$.
\end{definition}

\medskip
We are now ready to formulate the LP problem for the random hyperbolic system \eqref{RPDE1}-\eqref{RPDE2}.  
We denote the probability distribution function of the Young measure $\mathcal{V}_{t,x,\omega}$ by $f$ and set $F(x,t, \omega,\xi) = f(x,t,\omega,\xi) p(\omega)$,
thus,
\begin{align*}
& F: (0,T)  \times \Omega \times D \times \mathbb{R}^n  \to [0,\infty), \\
& \mbox{satisfying} \quad  \int_D \int_{\mathbb{R}^n} F(t,x,\omega, \xi)  d\xi d \omega  =1, \quad F(t, x, \omega, \xi) \geq 0, \quad (\omega, \xi) \in D\times \mathbb{R}^n.  
\end{align*}

Analogously to Section~\ref{sec:LP-Determin}, the linear optimization problem of entropy maximization for random nonlinear PDE \eqref{RPDE1}-\eqref{RPDE2} can be formulated as follows:
\begin{align}\label{R-LP1}
    \text{argmax}_{F(t, x, \omega, \xi)} \int_{\Omega} \int_D \int_{\mathbb{R}^n} \eta(\xi) F(t, x, \omega, \xi)  d \xi d\omega dx
\end{align}
subject to:
\begin{align}\label{R-LP2}
&  \int_D \int_{\mathbb{R}^n} F(t,x,\omega, \xi) d\xi d \omega  =1, \quad F(t, x, \omega, \xi) \geq 0,\\
         \label{R-LP3}
& \int_{\mathbb{R}^n} \left( \xi \frac{\partial F(t, x, \omega, \xi)}{\partial t}+f(\xi) \cdot \nabla_x F(t, x, \omega, \xi)\right) d \xi =0.
\end{align}

\subsection{Collocation framework and cost of quantum algorithms}\label{sec-SC}

In this section, we propose an approximation to the LP problem \eqref{R-LP1}-\eqref{R-LP3} 
by combining the finite volume method from Section~\ref{sec-det-dis} with the collocation method in the random space.
Let $(t, x, \xi) \in [0, T] \times \Omega \times [\xi_{\min}, \xi_{\max}]^n$ be divided into a uniform grid consisting of $N_t N_x^dN_{\xi}^n$ cells, as in the deterministic case described in Section~\ref{sec-det-dis}.  

Let $\{\omega_q\}_{q=1}^{N_{\omega}^m}$ be the collocation points with respect to $p(\omega)$. 

We seek 
\[
F_h = \frac{1}{h_{\xi}^n}  \sum_{j=1}^{N_t}\sum_{k}\sum_{q} \sum_{l}  F_{kql}^j  1_{K_t^j}1_{K_x^k}1_{\omega_q} 1_{K_{\xi}^l} 
\]
for the linear programming problem \eqref{R-LP1}-\eqref{R-LP3} as follows:
\begin{align}\label{RSC-LP1}
   &  \text{argmax}_F \sum_{k}\sum_{q} \sum_{l}  \Ov{\eta}_l F_{kql}^{j+1}, 
\end{align}
subject to 
\begin{align}\label{RSC-LP2}
& \sum_{q} \sum_{l}  F_{kql}^{j+1} =1, \qquad 
    F_{kql}^{j+1} \geq 0, \\  
    \label{RSC-LP3}
& \sum_{l} \left(\Ov{u}_l \frac{F_{kql}^{j+1}-F_{kql}^j}{\Delta t} + \sum_{i=1}^d \left[ (\Ov{f}_l)^{(i)} \frac{F_{k+e_i,ql}^j-F_{k-e_i,ql}^j}{2 h_x} - \varepsilon_{l} \Ov{u}_l \frac{F_{k+e_i,ql}^j- 2 F_{k,ql}^j +F_{k-e_i,ql}^j}{h_x^2}  \right] \right)=0.
\end{align}
As mentioned above, we apply the Lax-Friedrichs finite volume method for time-space discretization. 
\medskip

Analogous to the deterministic case in Section~\ref{sec-det-dis}, for  any fixed $j \in \{1, 2, \dots, N_T\},$, we assembly a vector $F^j$ of size $N_x^d N_{\omega}^m N_{\xi}^n$ with the entries $F_{kql}^j$:
\begin{enumerate}
\item 
\begin{align*}
    \sum_{k} \sum_{q} \sum_{l}  \Ov{\eta}_l F_{kql}^{j+1} = \mathbf{1}_{1\times N_x^dN_{\omega}^m} \otimes \eta_{N_{\xi}^n} \cdot F^{j+1} := \widetilde{\Xi} F^{j+1};
\end{align*}

\item 
\begin{align*}
    \sum_{q} \sum_{l} F_{kql}^{j+1} = \mathbb{I}_{N_x^d} \otimes \mathbf{1}_{1\times N_{\omega}^m N_{\xi}^n} \cdot F^{j+1} := \widetilde{A} F^{j+1};
\end{align*}

\item 
\begin{align*}
\mathbf{0}_{n N_x^d N_{\omega}^m}
= & \ \sum_{l} \Ov{u}_l F_{kql}^{j+1} + \sum_{l}\sum_{i=1}^d ( a_l^i F_{k-e_i,ql}^{j} + b_l^i F_{kql}^{j} + c_l^i F_{k+e_i,ql}^{j}) \\
= & \ \mathbb{I}_{N_x^dN_{\omega}^m} \otimes u_{N_{\xi}^n} \cdot F^{j+1} + \sum_{i=1}^d \Big( \mathbb{K}_{N_x^dN_{\omega}^m}^i(0)^T \otimes a_{N_{\xi}^n}  \\
& \qquad + \mathbb{I}_{N_x^dN_{\omega}^m} \otimes b_{N_{\xi}^n} + \mathbb{K}_{N_x^dN_{\omega}^m}^i(0) \otimes c_{N_{\xi}^n}  + \mathbb{R}^i_{n N_x^d N_{\omega}^m \times N_x^d N_{\omega}^m N_{\xi}^n} \Big) F^j \\
:= & \  \widetilde{B} F^{j+1} + \widetilde{D} F^{j}.
\end{align*}
\end{enumerate}
Consequently, the LP problem \eqref{RSC-LP1}-\eqref{RSC-LP3}  can be written  in the following matrix form: 
\begin{align}
& \text{argmax}_{\mathbf{F}} \, \Xi\, \mathbf{F}, \nonumber \\
& \mathbf{F} \geq \mathbf{0}_{N_tN_x^dN_{\omega}^mN_{\xi}^n \times 1}, \quad M\mathbf{F}= (\mathbf{1}_{N_tN_x^d \times 1}; \mathbf{0}_{n N_tN_x^dN_{\omega}^m \times 1}), 
\end{align}
with
\begin{align*}
& {M}=\begin{pmatrix}
            {A} \\
            {B}
\end{pmatrix}, \quad {\Xi} = \mathbf{1}_{1\times N_t} \otimes \widetilde{\Xi}, \quad 
{A} = \mathbb{I}_{N_t} \otimes \widetilde{A}, \quad 
{B} = \mathbb{J}_{N_t}(0)^T \otimes \widetilde{D} +  \mathbb{I}_{N_t} \otimes \widetilde{B}.
\end{align*}
In Table~\ref{tab:RSC} we have computed the sizes of the relevant properties of $M$. \\

\begin{table}[h] 
\centering
\caption{Sizes of parameters related to the LP problem \eqref{RSC-LP1}-\eqref{RSC-LP3}. Here, $d$, $n$ and $m$ are much smaller than $N_t, N_x,N_{\omega},N_{\xi}$. Here the cost is in $\tilde{O}$ since we neglect multiplicative factors of $d, m, n$.} \label{tab:RSC} 
\begin{tabular}{|l|l|ll} 
\cline{1-2}
Parameter  &  LP-FxM $\tilde{O}(\cdot)$  &  \\ \cline{1-2}
$r$ & $N_tN_x^d N_{\omega}^m$ &  \\ \cline{1-2}
$s$ & $r N_{\xi}^n = N_t N_x^d N_{\omega}^m N_{\xi}^n$  &   \\ \cline{1-2}
$R_1$ & $N_t N_x^d$  &  \\ \cline{1-2}
$\text{nnz}(M)$  & $N_t N_x^dN_{\omega}^m N_{\xi}^n$  &  \\ \cline{1-2}
$s_M$ & $N_{\omega}^m N_{\xi}^n$  &\\ \cline{1-2}
\end{tabular}
\end{table}

Inserting the results of Table~\ref{tab:RSC} into Table~\ref{tab:1}, we obtain the costs for outputting classical $\mathbf{F}$ in Table~\ref{tab:UQcost1}. Similarly to  Section~\ref{sec:youngnonlinearpde1}, we do not expect quantum advantage in $N_t$ or $N_x$. However, there is the possibility of quantum advantage in both $N_{\xi}$ and $N_{\omega}$. In fact, we can see that using QCP (gate complexity), compared to IPM, the quantum advantage is $\tilde{O}(N_{\omega}^{m}N_{\xi}^n\epsilon)$. Since $n$ corresponds to the number of equations in the PDE system, this number is typically not so high, so the advantage can become marginal when $n$ is small. Here, $m$ denotes the dimension of the random space, and it can be very large even when $n$ is small. For example, in the case of first-order methods, if we choose $N_{\xi} \sim N_{\omega} \sim N \sim 1/\epsilon$, then the quantum advantage scales as $\tilde{O}(N^{m+n-1})$. Thus, we observe a polynomial quantum advantage in recovering the Young measure with respect to both $n$ and $m$.  
\begin{table}[h] 
\centering
\caption{Comparison of costs to output classical $\mathbf{F}$ (using Table~\ref{tab:RSC}) different algorithms in terms of $q=N_tN_x^d$, $m, n$ and discretisation sizes $N_{\omega}, N_{\xi}$. 
} \label{tab:UQcost1}
\begin{tabular}{|l|l|l|ll} 
\cline{1-3}
Algorithm  &  Time/query complexity  & Comments  &     \\ \cline{1-3}
IPM \cite{lee2015efficient} & $\tilde{O}(q^{2.5}N_{\omega}^{5m/2}N_{\xi}^{5n/2})$ & $\log(1/\epsilon)$ dependence  &     \\ \cline{1-3}
SIPM \cite{cohen2021solving} & $\tilde{O}(q^{2.38}N_{\omega}^{2.38m}N_{\xi}^{2.38n})$ &  $\log(1/\epsilon)$ dependence  &     \\ \cline{1-3}
QIPM\cite{kerenidis2020quantum} & $ \tilde{O}(\kappa^3_{Newt} q^{2} N_{\omega}^{2m}N_{\xi}^{2n}/\delta^2)$  & $\kappa_{Newt}$ bound to be determined &   \\ \cline{1-3}
QZSG\cite{van2019quantum} & $\tilde{O}((q/\epsilon)^{3.5}N_{\omega}^{m/2}N_{\xi}^{n/2})$ & Assuming $\inf \|\lambda\|_1\sim \tilde{O}(1)$ &     \\ \cline{1-3}
QCP \cite{augustino2023quantum}& $\tilde{O}(q^{1.5}N_{\omega}^{m/2} N_{\xi}^{n/2}/\epsilon)$ & Gate complexity  $\tilde{O}(q^{2.5}N_{\omega}^{3m/2}N_{\xi}^{3n/2}/\epsilon)$&    \\ \cline{1-3}
\end{tabular}
\end{table}

An important distinction between deterministic and random nonlinear PDEs is the computational cost of the classical algorithms that \textit{directly} solve the original PDEs rather than their Young measures. In the deterministic nonlinear PDE case discussed in the previous sections, we saw that the cost of classical algorithms that directly solve the original PDEs (instead of the Young measures) is $O(dN_t N_x^d)$, which is always less than the cost of any classical or quantum algorithm for computing the Young measure $\mathbf{F}$. The same conclusion does not necessarily hold for random nonlinear PDEs. If one uses the stochastic collocation method to directly recover the solution of a system of $n$ nonlinear PDEs in $d$ spatial dimensions and $m$ random dimensions, the cost is $O(nN_tN_x^d N_{\omega}^m)=\tilde{O}(qN_{\omega}^m)$. We already see from Table~\ref{tab:UQcost1} that both QZSG and QCP have query complexities that scale with $N_{\omega}$ like $N_{\omega}^{m/2}$. This suggests that if $m$ is large enough, then it costs less for the quantum algorithm to recover $\mathbf{F}$. If one uses the query complexity of QCP and assuming first-order methods where $1/\epsilon=N \sim N_t \sim N_x \sim N_{\omega}\sim N_{\xi}$, then comparing the QCP query complexity  $\tilde{O}(q^{1.5}NN_{\omega}^{m/2}N_{\xi}^{n/2})=\tilde{O}(N^{(m+n)/2+1.5d+2.5})$ (to obtain $\mathbf{F}$) with the direct classical solver $\tilde{O}(qN_{\omega}^m)=\tilde{O}(N^{m+d+1})$, we can see that there is a quantum advantage in $N$  as long as $m>n+d+3$. \\

This advantage is important to emphasize, since $\mathbf{F}$ in fact has much more information about the properties of the PDE  (especially in highly nonlinear settings) than a solution $u$ computed by standard numerical methods that directly solve the nonlinear PDE. \\

Using first order methods for simplicity $N=1/\epsilon \sim N_t \sim N_x\sim N_{\xi} \sim N_{\omega}$, one can see from Table~\ref{tab:uncertaindirect} that for our three prototype nonlinear PDEs, for only moderate $m$, quantum algorithms can recover $\mathbf{F}$ more efficiently -- by polynomial factors of $N_{\omega}$ -- compared to direct classical solvers of the random nonlinear PDE. For example, for the compressible Euler or Navier-Stokes equations, one can see that the quantum cost is more efficient by a factor of $\sim N^{m/2-5.5}$. In real applications, $m$ can even be as high as $100$, thus this could lead to a polynomial factor of $N^{44.5}$.

\begin{table}[h] 
\centering
\caption{Comparison of quantum cost to output classical $\mathbf{F}$ and classical cost to output solution $u$ by direct PDE numerical solver (stochastic collocation). We use first-order methods to compare the QCP query complexity $\tilde{O}(N^{(m+n)/2+1.5d+2.5})$ to the direct classical stochastic collocation cost $\tilde{O}(N^{m+d+1})$ and $d=3$.
} \label{tab:uncertaindirect}
\begin{tabular}{|l|l|l|ll} 
\cline{1-3}
Equation &  Quantum query complexity ($\mathbf{F}$) & Cost of direct classical solver ($u$) &     \\ \cline{1-3}
Burgers; Allen-Cahn & $\tilde{O}(N^{m/2+7.5})$ & $\tilde{O}(N^{m+4})$  &     \\ \cline{1-3}
Barotropic Euler &  $\tilde{O}(N^{m/2+9})$&  $\tilde{O}(N^{m+4})$   &     \\ \cline{1-3}
Full Euler; NSF & $\tilde{O}(N^{m/2+9.5})$  & $\tilde{O}(N^{m+4})$  &   \\ \cline{1-3}
\end{tabular}
\end{table}

However, we note that this advantage is \textit{not} the same as comparing the directly obtained solution $u$ with $\bar{u}=\int_D \int_{\Omega} \xi F(t, x, \omega, \xi) d \xi d\omega$, where the latter is the expected value (observable) with respect to the Young measure. This is because the cost in performing the numerical integration over $\xi$ and $\omega$ is already of size $\tilde{O}(N_{\xi}^nN_{\omega}^m)$, which carries its own curse of dimensionality problem. Here we do not assume any quantum advantage in the integration step since the output of quantum LP solvers considered here is the classical $\mathbf{F}$. However, if quantum LP algorithms are developed that output instead $|F\rangle$ or $|\sqrt{F}\rangle$, then there could be a possibility of overall quantum advantage in obtaining the measure-valued solutions themselves, as briefly explained in Section~\ref{sec:youngnonlinearpde1}.\\

We note that Monte Carlo methods for this integration step are generally not preferred when an accurate approximation is desired. However, if one were to use Monte Carlo methods for this integration step (and thereby bypassing the curse of dimensionality here), then this would imply that one ought to also use Monte Carlo based methods for the direct classical solver (instead of using stochastic collocation methods), in which case there would also be no quantum advantage since the cost of classical Monte Carlo solvers for the random PDEs will not scale with $N_{\omega}^m$ but with exponentially less dependence on $m$.

\section{Outlook and open questions} \label{sec:outlook}

There are two goals in developing quantum algorithms for computing the Young measures of associated nonlinear PDEs. 
First,  there are many scenarios where computing the Young measures 
is of strong scientific interest. This is due to the fact that the measure-valued formulation describes physically relevant solutions for problems that contain singularities, ill-posedness, or uncertainties, and for problems in which one is not interested in the solutions of the underlying PDEs at each grid point, but rather their probability density distributions. The latter arises in many physical or engineering applications, including combustion and turbulence modelling \cite{Lim-Glimm, klingenberg2025, Feireisl-Lukacova, Chertock2026, FLSY2026}. In these cases, one should compare the costs of quantum LP algorithms with those of  classical LP, in which we have demonstrated that some quantum algorithms, for example, the quantum central path algorithms \cite{augustino2023quantum}, offer polynomial advantages. 

The second goal is to investigate where the LP formulation, due to its linearity, is able to provide greater efficiency when using quantum algorithms for the original nonlinear PDEs. For uncertain PDE problems, when the number of random uncertain variables is sufficient large, the cost saving can be significant. But for deterministic problems, our analysis does not find any quantum advantages over the direct classical solvers for the original PDEs.  We believe that there is room for further improvement, and hopefully, this program will eventually lead to quantum algorithms with quantum advantages. Here we provide a few possible open questions:

\begin{itemize}

\item 
Utilize the special structure of the equations and solutions to reduce the computational costs of the LP problems. One possibility is to utilize the special feature of the Young measure, which is close to a Dirac mass, thus it is highly concentrated around the solutions of the original PDEs in physical space. One may take advantage of this special structure, or sparsity, to significantly reduce computational cost by using adaptive (discretization) meshes concentrated over the support of the Dirac measure.

\item Using the Lasserre hierarchy, which evolves the moments of the Young measure \cite{cardoen2024} using Semi-Definite Programming (SDP). (See also the discussion in Section \ref{Sec:cost} for why quantum algorithms for SDP are more suited to a direct SDP problem than for an LP problem).
Is it possible to obtain  more efficient algorithms than those using the Young measure based LP formulation?

\item Improved quantum LP algorithms and also new quantum LP algorithms that output the solution as a quantum state (like $|F\rangle$ or $|\sqrt{F}\rangle$) instead of as a classical state.

\end{itemize}

\section*{Acknowledgements}
 
S.J. and N.L. acknowledge the support of the NSFC grant No. 12341104, the Shanghai Pilot Program for Basic Research,  the Science and Technology Commission of Shanghai Municipality (STCSM) grant no. 24LZ1401200 (21JC1402900), the Shanghai Jiao Tong University 2030 Initiative, and the Fundamental Research Funds for the Central Universities. N.L. is also supported by grant NSFC No. 12471411. 

M.L.-M. gratefully acknowledges the support of DFG Project 5258 3336 funded within the Focused Program
SPP 2410 ``Hyperbolic Balance Laws: Complexity, Scales and Randomness'' and of the Mainz
Institute of Multiscale Modeling. She is also grateful to the Gutenberg Research College for supporting her research.

Y.Y. acknowledges the support of the National Natural Science Foundation of China under grant No.\ 12401527, and the Natural Science Foundation of Jiangsu Province under grant No. BK20241364.

\bibliographystyle{siam}

\bibliography{references}

\begin{thebibliography}{10}

\bibitem{ALL2023LCH}
{\sc D.~An, J.~Liu, and L.~Lin}, {\em Linear combination of {H}amiltonian simulation for non-unitary dynamics with optimal state preparation cost}, Phys. Rev. Lett., 131 (2023), p.~150603.

\bibitem{apers2023quantum}
{\sc S.~Apers and S.~Gribling}, {\em Quantum speedups for linear programming via interior point methods}, arXiv preprint arXiv:2311.03215,  (2023).

\bibitem{augustino2023quantum}
{\sc B.~Augustino, J.~Leng, G.~Nannicini, T.~Terlaky, and X.~Wu}, {\em A quantum central path algorithm for linear optimization}, arXiv preprint arXiv:2311.03977,  (2023).

\bibitem{BerryChilds2017ODE}
{\sc D.~W. Berry, A.~M. Childs, A.~Ostrander, and G.~Wang}, {\em Quantum algorithm for linear differential equations with exponentially improved dependence on precision}, Commun. Math. Phys., 356 (2017), pp.~1057--1081.

\bibitem{brandao2017quantum}
{\sc F.~G. Brandao and K.~M. Svore}, {\em Quantum speed-ups for solving semidefinite programs}, in 2017 IEEE 58th Annual Symposium on Foundations of Computer Science (FOCS), IEEE, 2017, pp.~415--426.

\bibitem{Breit}
{\sc D.~Breit, E.~Feireisl, and M.~Hofmanov\'{a}}, {\em Dissipative solutions and semiflow selection for the complete {E}uler system}, Comm. Math. Phys., 376 (2020), pp.~1471--1497.

\bibitem{MR1816648}
{\sc A.~Bressan}, {\em Hyperbolic systems of conservation laws}, vol.~20 of Oxford Lecture Series in Mathematics and its Applications, Oxford University Press, Oxford, 2000.
\newblock The one-dimensional Cauchy problem.

\bibitem{BEHL}
{\sc A.~Brunk, H.~Egger, O.~Habrich, and M.~Luk\'{a}\v{c}ov\'{a}-Medvi\v{d}ov\'{a}}, {\em Stability and discretization error analysis for the {Cahn-Hilliard} system via relative energy estimates}, ESIAM: M2AN, 57 (2023), pp.~1297--1322.

\bibitem{cardoen2024}
{\sc C.~Cardoen, S.~Marx, A.~Nouy, and N.~Seguin}, {\em A moment approach for entropy solutions of parameter-dependent hyperbolic conservation laws}, Numer. Math., 156 (2024), pp.~1289--1324.

\bibitem{casares2020quantum}
{\sc P.~A. Casares and M.~A. Martin-Delgado}, {\em A quantum interior-point predictor--corrector algorithm for linear programming}, Journal of physics A: Mathematical and Theoretical, 53 (2020), p.~445305.

\bibitem{Chertock2026}
{\sc A.~Chertock, M.~Herty, A.~I. A.~Ishakov, A.~Kurganov, and M.~Luk\'a\v{c}ov\'a-Medvi\v{d}ov\'a}, {\em Numerical study of random {Kelvin-Helmholtz} instability}, Comm. Comp. Phys.,  (2026).

\bibitem{Chiodaroli-DeLellis-Kreml:2015}
{\sc E.~Chiodaroli, C.~De~Lellis, and O.~Kreml}, {\em Global ill-posedness of the isentropic system of gas dynamics}, Commun. Pure Appl. Math., 68 (2015), pp.~1157--1190.

\bibitem{CF}
{\sc E.~Chiodaroli and E.~Feireisl}, {\em On the density of ``wild'' initial data for the barotropic {E}uler system}, Ann. Mat. Pura Appl. (4), 203 (2024), pp.~1809--1817.

\bibitem{CF_21}
{\sc E.~Chiodaroli, O.~Kreml, V.~M{\'a}cha, and S.~Schwarzacher}, {\em Non-uniqueness of admissible weak solutions to the compressible {Euler} equations with smooth initial data}, Trans. Am. Math. Soc., 374 (2021), pp.~2269--2295.

\bibitem{chu2025solving}
{\sc S.~Chu, M.~Herty, M.~Luk\'a\v{c}ov\'a-Medvi\v{d}ov\'a, and Y.~Zhou}, {\em Solving random hyperbolic conservation laws using linear programming}, 2025.

\bibitem{cohen2021solving}
{\sc M.~B. Cohen, Y.~T. Lee, and Z.~Song}, {\em Solving linear programs in the current matrix multiplication time}, Journal of the ACM (JACM), 68 (2021), pp.~1--39.

\bibitem{MR3468916}
{\sc C.~M. Dafermos}, {\em Hyperbolic conservation laws in continuum physics}, vol.~325 of Grundlehren der mathematischen Wissenschaften [Fundamental Principles of Mathematical Sciences], Springer-Verlag, Berlin, fourth~ed., 2016.

\bibitem{DLSz}
{\sc C.~De~Lellis and L.~Sz\'{e}kelyhidi, Jr.}, {\em The {E}uler equations as a differential inclusion}, Ann. of Math. (2), 170 (2009), pp.~1417--1436.

\bibitem{DlSz1}
\leavevmode\vrule height 2pt depth -1.6pt width 23pt, {\em On admissibility criteria for weak solutions of the {E}uler equations}, Arch. Ration. Mech. Anal., 195 (2010), pp.~225--260.

\bibitem{DeLellis-Szekelyhidi:2009}
{\sc C.~De~Lellis and L.~Sz\'{e}kelyhidi~Jr.}, {\em The {Euler} equations as a differential inclusion}, Ann. Math., 170 (2009), pp.~1417--1436.

\bibitem{DeTa:2011}
{\sc P.~Degond and M.~Tang.}, {\em All speed scheme for the low {Mach} number limit of the isentropic {Euler} equations}, Commun. Comput. Phys., 10 (2011), pp.~1--31.

\bibitem{DiPerna-Majda:1987}
{\sc R.~DiPerna and A.~Majda}, {\em {Oscillations and concentrations in weak solutions of the incompressible fluid equations}}, Comm. Math. Phys., 108 (1987), pp.~667--689.

\bibitem{DiPerna:1983a}
{\sc R.~J. DiPerna}, {\em Generalized solutions to conservation laws}, Springer Netherlands, 1983, pp.~305--309.

\bibitem{DiPerna1985}
{\sc R.~J. DiPerna}, {\em Measure-valued solutions to conservation laws}, Arch. Rational Mech. Anal., 88 (1985), pp.~223--270.

\bibitem{MR3409135}
{\sc L.~C. Evans and R.~F. Gariepy}, {\em Measure theory and fine properties of functions}, Textbooks in Mathematics, CRC Press, Boca Raton, FL, revised~ed., 2015.

\bibitem{MR257325}
{\sc H.~Federer}, {\em Geometric measure theory}, vol.~Band 153 of Die Grundlehren der mathematischen Wissenschaften, Springer-Verlag New York, Inc., New York, 1969.

\bibitem{Feireisl-Jungel-Lukacova}
{\sc E.~Feireisl, A.~J\"ungel, and M.~Luk\'a\v{c}ov\'a-Medvi\v{d}ov\'a}, {\em Maximal dissipation and well-posedness of the {Euler} system of gas dynamics}, 2025.

\bibitem{FL_18}
{\sc E.~Feireisl and M.~Luk{\'a}{\v{c}}ov{\'a}-Medvi\v{d}ov{\'a}}, {\em Convergence of a mixed finite element-finite volume scheme for the isentropic {Navier}-{Stokes} system via dissipative measure-valued solutions}, Found. Comput. Math., 18 (2018), pp.~703--730.

\bibitem{FLSS_22}
{\sc E.~Feireisl, M.~Luk{\'a}{\v{c}}ov{\'a}-Medvi\v{d}ov{\'a}, S.~Schneider, and B.~She}, {\em Approximating viscosity solutions of the {Euler} system}, Math. Comput., 91 (2022), pp.~2129--2164.

\bibitem{MR4674068}
{\sc E.~Feireisl and M.~Luk\'a\v{c}ov\'a-Medvi\v{d}ov\'a}, {\em Convergence of a stochastic collocation finite volume method for the compressible {N}avier-{S}tokes system}, Ann. Appl. Probab., 33 (2023), pp.~4936--4963.

\bibitem{Feireisl-Lukacova}
{\sc E.~Feireisl and M.~Luk\'{a}\v{c}ov\'{a}-Medvi\v{d}ov\'{a}}, {\em Well-posedness of the {E}uler system of gas dynamics}, 2025.

\bibitem{MR4130543}
{\sc E.~Feireisl, M.~Luk\'a\v{c}ov\'a-Medvi\v{d}ov\'a, and H.~Mizerov\'a}, {\em Convergence of finite volume schemes for the {E}uler equations via dissipative measure-valued solutions}, Found. Comput. Math., 20 (2020), pp.~923--966.

\bibitem{MR4390192}
{\sc E.~Feireisl, M.~Luk\'a\v{c}ov\'a-Medvi\v{d}ov\'a, H.~Mizerova, and B.~She}, {\em Numerical analysis of compressible fluid flows}, vol.~20 of MS\&A. Modeling, Simulation and Applications, Springer, Cham, 2021.

\bibitem{FLSY2026}
{\sc E.~Feireisl, M.~Luk\'a\v{c}ov\'a-Medvi\v{d}ov\'a, B.~She, and Y.~Yuan}, {\em Temperature-driven turbulence in compressible fluid flows}, 2026.

\bibitem{feireisl-lukacova2025}
{\sc E.~Feireisl, M.~Luk\'a\v{c}ov\'a-Medvi\v{d}ov\'a, and C.~Yu}, {\em Oscillatory approximations and maximum entropy principle for the {E}uler system of gas dynamics}, 2025.

\bibitem{fjordholm2016computation}
{\sc U.~S. Fjordholm, S.~Mishra, and E.~Tadmor}, {\em On the computation of measure-valued solutions}, Acta numerica, 25 (2016), pp.~567--679.

\bibitem{HHL2009}
{\sc A.~W. Harrow, A.~Hassidim, and S.~Lloyd}, {\em Quantum algorithm for linear systems of equations}, Phys. Rev. Lett., 103 (2009), pp.~150502, 4 pp.

\bibitem{2023analogPDE}
{\sc S.~Jin and N.~Liu}, {\em Analog quantum simulation of partial differential equations}, Quantum Science and Technology,  (2023).

\bibitem{JinLiu2022nonlinear}
{\sc S.~Jin and N.~Liu}, {\em Quantum algorithms for nonlinear partial differential equations}, Bull. Sci. math, 194 (2024), p.~103457.

\bibitem{hjviscosityarxiv}
{\sc S.~Jin and N.~Liu}, {\em Quantum algorithms for viscosity solutions to nonlinear hamilton-jacobi equations based on an entropy penalisation method}, arXiv preprint arXiv:2512.07919,  (2025).

\bibitem{schrpra}
{\sc S.~Jin, N.~Liu, and Y.~Yu}, {\em Quantum simulation of partial differential equations: Applications and detailed analysis}, Physical Review A, 108 (2023), p.~032603.

\bibitem{schrprl}
\leavevmode\vrule height 2pt depth -1.6pt width 23pt, {\em Quantum simulation of partial differential equations via schr{\"o}dingerization}, Physical Review Letters, 133 (2024), p.~230602.

\bibitem{kerenidis2020quantum}
{\sc I.~Kerenidis and A.~Prakash}, {\em A quantum interior point method for lps and sdps}, ACM Transactions on Quantum Computing, 1 (2020), pp.~1--32.

\bibitem{klingenberg2025}
{\sc C.~Klingenberg, S.~Markfelder, and E.~Wiedemann}, {\em Maximal turbulence as a selection criterion for measure-valued solutions}, 2025.

\bibitem{lee2015efficient}
{\sc Y.~T. Lee and A.~Sidford}, {\em Efficient inverse maintenance and faster algorithms for linear programming}, in 2015 IEEE 56th annual symposium on foundations of computer science, IEEE, 2015, pp.~230--249.

\bibitem{Lim-Glimm}
{\sc H.~Lim, Y.~Yu, J.~Glimm, X.~Li, and D.~Sharp}, {\em Chaos, transport and mesh convergence for fluid mixing.}, Acta Math. Appl. Sin. Engl., 24 (2008).

\bibitem{Liu-PNAS}
{\sc J.-P. Liu, H.~O.~i. Kolden, H.~K. Krovi, N.~F. Loureiro, K.~Trivisa, and A.~M. Childs}, {\em Efficient quantum algorithm for dissipative nonlinear differential equations}, Proc. Natl. Acad. Sci. USA, 118 (2021), pp.~Paper No. e2026805118, 6.

\bibitem{liu2025quantum}
{\sc N.~Liu, M.~Minervini, D.~Patel, and M.~M. Wilde}, {\em Quantum thermodynamics and semi-definite optimization}, arXiv preprint arXiv:2505.04514,  (2025).

\bibitem{Klingenberg1}
{\sc S.~Markfelder and C.~Klingenberg}, {\em The {R}iemann problem for the multidimensional isentropic system of gas dynamics is ill-posed if it contains a shock}, Arch. Ration. Mech. Anal., 227 (2018), pp.~967--994.

\bibitem{MR3793404}
{\sc S.~Mishra and C.~Schwab}, {\em Monte-{C}arlo finite-volume methods in uncertainty quantification for hyperbolic conservation laws}, in Uncertainty quantification for hyperbolic and kinetic equations, vol.~14 of SEMA SIMAI Springer Ser., Springer, Cham, 2017, pp.~231--277.

\bibitem{qipmreview}
{\sc M.~Mohammadisiahroudi, Z.~Wu, P.~Sampourmahani, A.~Harkness, and T.~Terlaky}, {\em Quantum interior point methods: A review of developments and an optimally scaling framework}, arXiv preprint arXiv:2512.06224,  (2025).

\bibitem{van2019quantum}
{\sc J.~van Apeldoorn and A.~Gily{\'e}n}, {\em Quantum algorithms for zero-sum games}, arXiv preprint arXiv:1904.03180,  (2019).

\end{thebibliography}

\appendix

\section{Numerical simulations}
\label{app:simulations}
In the following three subsections, we justify the mathematical validity of the LP formulations \eqref{LP1}-\eqref{LP3} on three prototype nonlinear PDEs: the inviscid Burgers equation, the barotropic Euler equations, and the Allen-Cahn equation.

\subsection{The Inviscid  Burgers equation} \label{sec:burgers}

Let us define the solution $u$, the energy $\hat{E}$ and the energy defect $\mathfrak{E}$ as follows
\begin{align}
u = \int_{\mathbb{R}} \xi F(t, x, \xi) d \xi, \quad 
\hat{E} = \int_{\mathbb{R}} \xi^2 F(t, x, \xi) d \xi, \quad  \mathfrak{E} = \hat{E} - u^2 .
\end{align}

\begin{figure}[htbp]
	\setlength{\abovecaptionskip}{0.cm}
	\setlength{\belowcaptionskip}{-0.cm}
	\centering

 	\begin{subfigure}{0.5\textwidth}
        \centering
        \includegraphics[width=\linewidth]{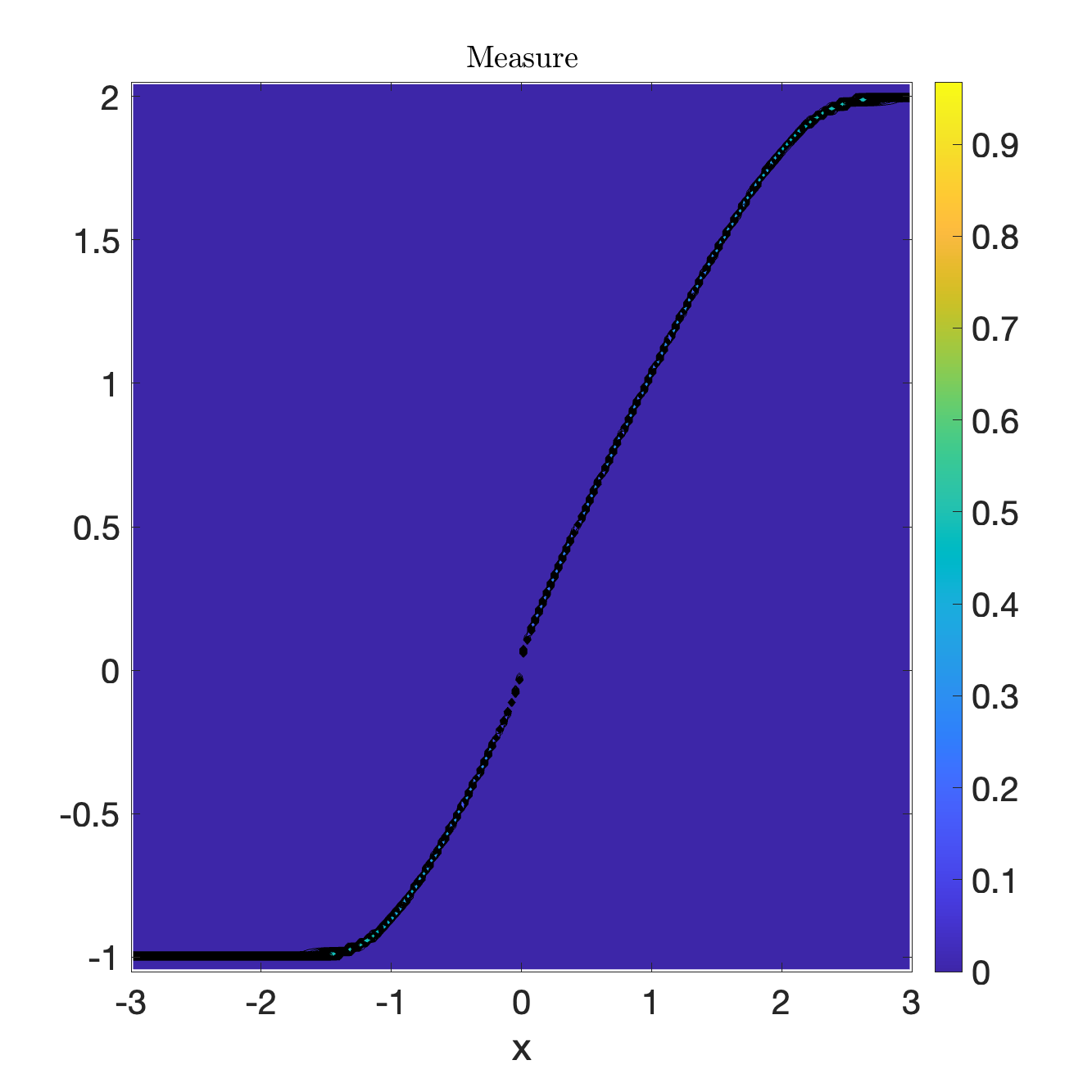}
    	\end{subfigure}
	\begin{subfigure}{0.48\textwidth}
        \centering
        \includegraphics[width=\textwidth]{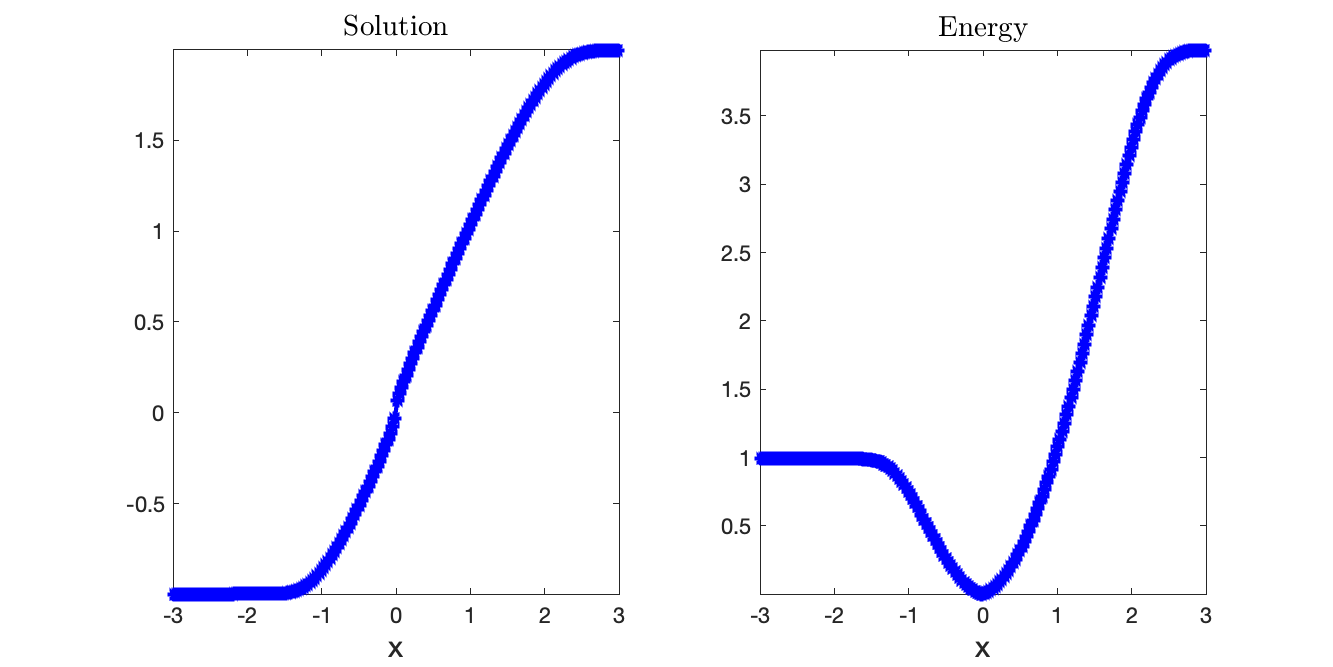}\\
       \includegraphics[width=\textwidth]{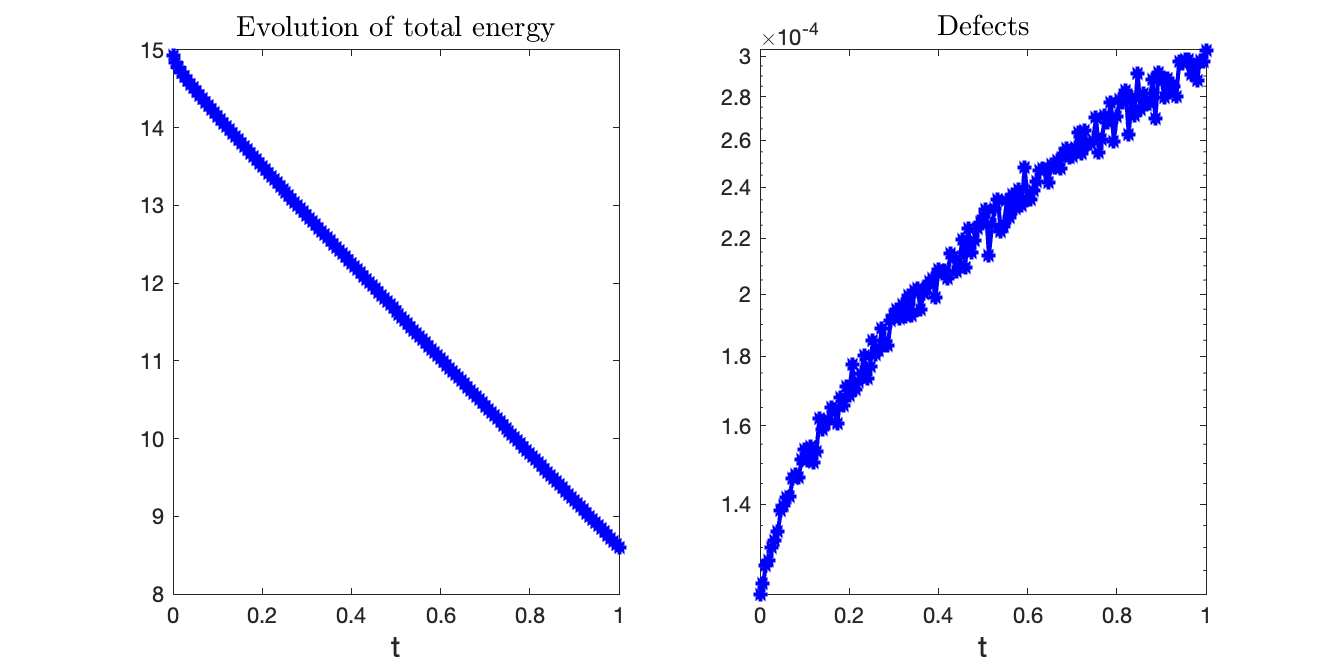}
    	\end{subfigure}
	\caption{  \small{Experiment 1 for the inviscid Burgers equation: the measure characterized by $F(T,x,\xi)$, solution $u(T,x)$, energy $\hat{E}(T,x)$, the evolution of total energy $ \int_{\Omega}\hat{E} dx$ and the total energy defects $\int_{\Omega}\mathfrak{E} dx $ with $T=1$ and $(N_t,N_x,N_{\xi}) = (150,200,200)$.}}\label{ex1-fig-Burgers}
\end{figure}

\begin{example}[Rarefaction wave]\label{example:1D-rarefaction}\rm
	The computational domain is $[-3,3]$, and the outflow boundary condition is applied. 
	The initial data is 
	\begin{equation*}
		u_0(x) = \begin{cases}
		-1,& \mbox{if}~ x < 0,\\
		2,& \mbox{if}~ x > 0.
		\end{cases}
	\end{equation*} 
In the experiment we take $T=1$ and $[\xi_{\min}, \xi_{\max}] = [-1.05, 2.05]$.  

Figure~\ref{ex1-fig-Burgers} shows the measure characterized by $F(T,x,\xi)$, i.e.\ $\int_{K_{\xi}^l} F(T,x,\xi) d \xi $ or $F_{kl}^{N_t}$, solution $u(T,x)$, energy $\hat{E}(T,x)$, the evolution of total energy $ \int_{\Omega}\hat{E} dx$ and the total energy defects $\int_{\Omega}\mathfrak{E} dx $ obtained with the discretization parameters $(N_t,N_x,N_{\xi}) = (150,200,200)$.  Further, the $L^p$-errors and corresponding rates on  consecutively refined meshes are presented in Table~\ref{table1}. 
\end{example}

\begin{table}[htbp]
	\centering
	\caption{Experiment 1  for the  inviscid Burgers equation (rarefaction wave): error and convergence rate.  } \label{table1}
	\begin{tabular}{|c|cccc|}
		\hline
		$(N_t,N_x,N_f)$ & $L^1$-error &  rate & $L^2$-error  & rate \\
		\hline
		(10, 15, 10) & 2.3182e-01&  -	&2.7073e-01&	- \\
		(20, 30, 20) & 1.2837e-01&  0.8527  	&1.5941e-01&	0.7641 \\
		(40, 60, 40) & 7.9106e-02&  0.6984 	&1.0612e-01&	0.5870 \\
		(80, 120, 80) & 5.6584e-02&  0.4834	&7.7059e-02&	0.4617 \\
		(160, 240, 160) & 3.3934e-02&   0.7377	&5.0642e-02&	0.6056 \\
		\hline
	\end{tabular}
\end{table}

\begin{figure}[htbp]
	\setlength{\abovecaptionskip}{0.cm}
	\setlength{\belowcaptionskip}{-0.cm}
	\centering
 	\begin{subfigure}{0.5\textwidth}
        \centering
        \includegraphics[width=\linewidth]{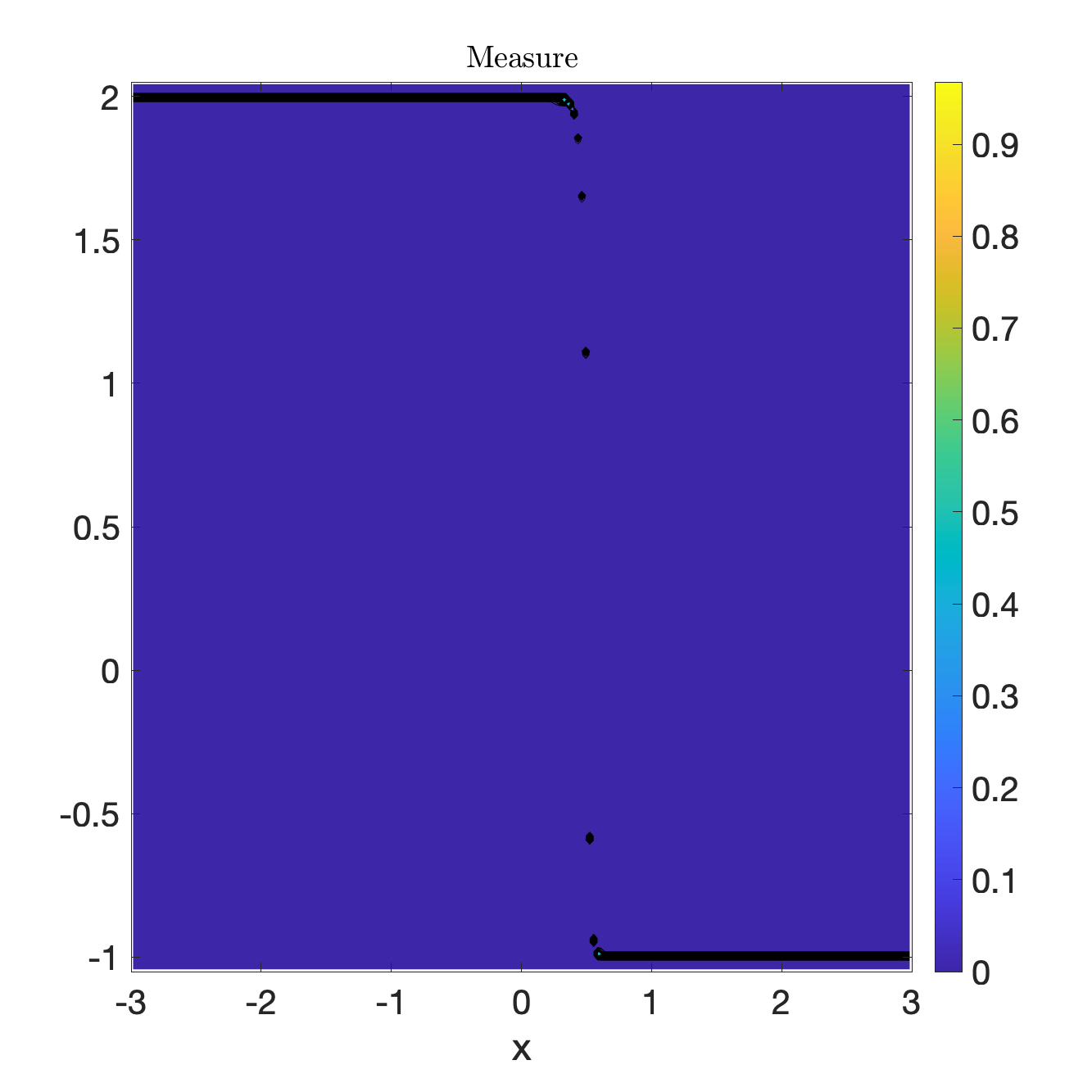}
    	\end{subfigure}
	\begin{subfigure}{0.48\textwidth}
        \centering
        \includegraphics[width=\textwidth]{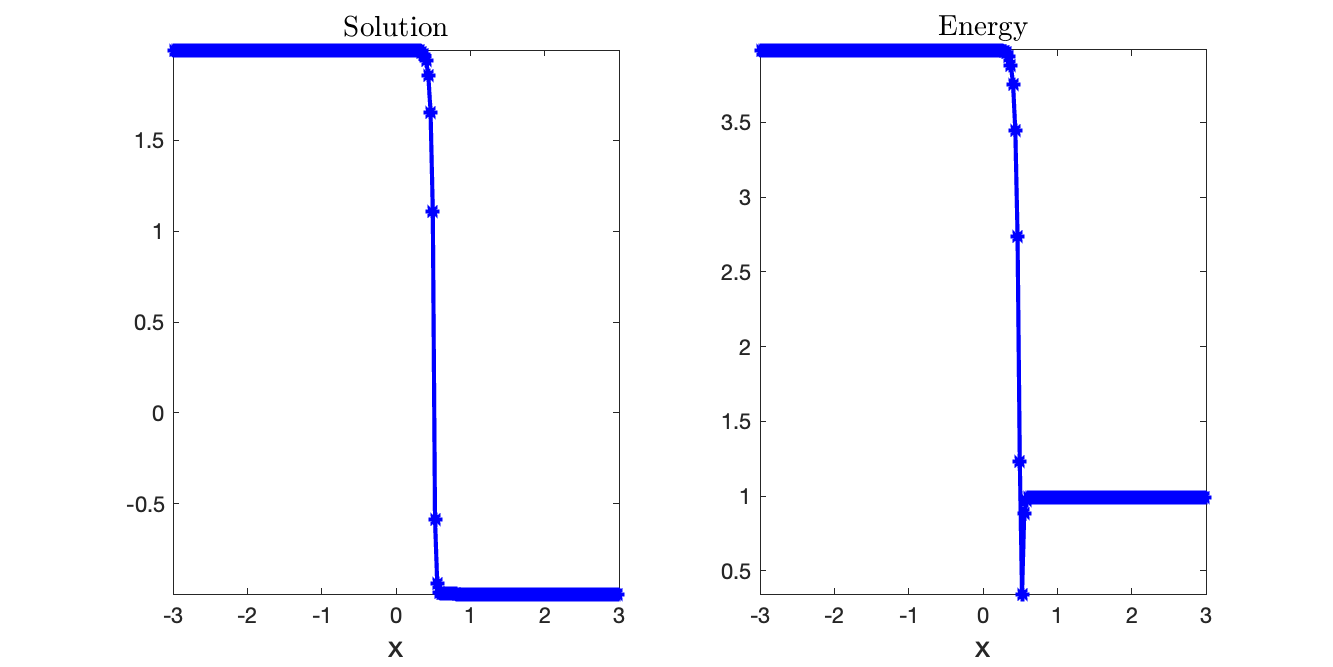}\\
       \includegraphics[width=\textwidth]{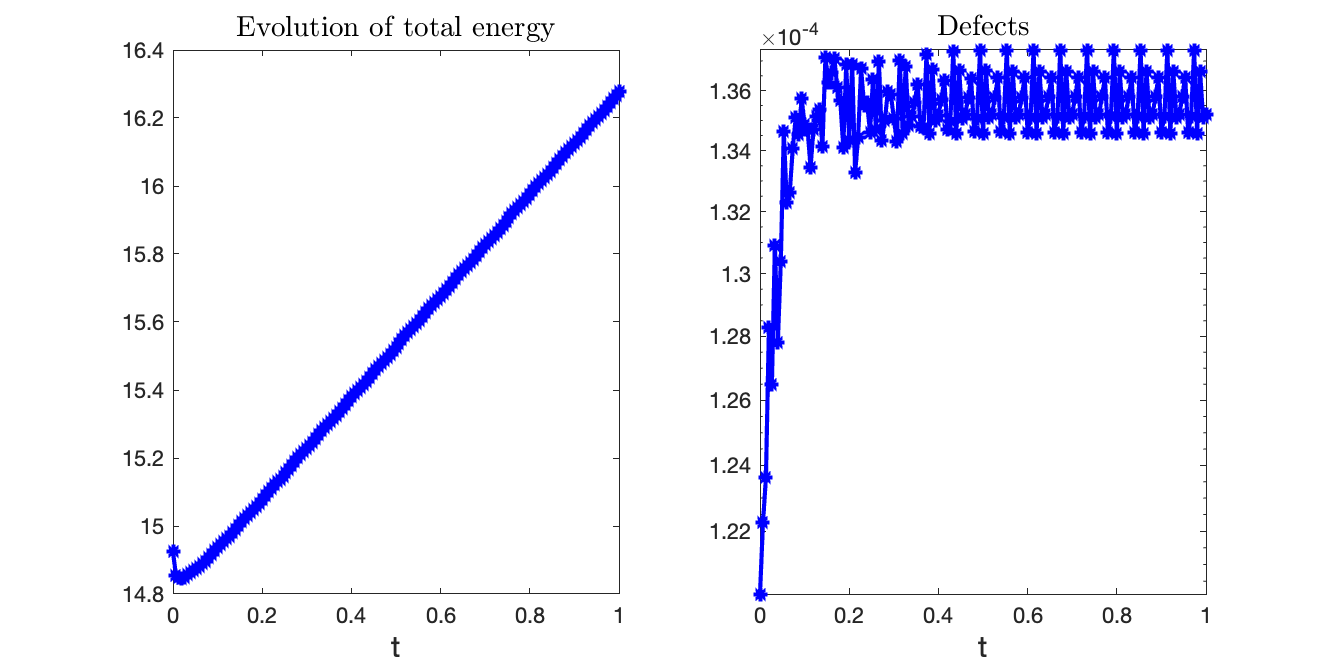}
    	\end{subfigure}
	\caption{  \small{Experiment 2 for the inviscid Burgers equation: the measure characterized by $F(T,x,\xi)$, solution $u(T,x)$, energy $\hat{E}(T,x)$, the evolution of total energy $ \int_{\Omega}\hat{E} dx$ and the total energy defects $\int_{\Omega}\mathfrak{E} dx $ with $T=1$ and $(N_t,N_x,N_f) = (150,200,200)$.}}\label{ex2-fig-Burgers}
\end{figure}

\begin{example}[Shock wave]\label{example:1D-shock}\rm
	The computational domain is $[-3,3]$. 
	The initial data is 
	\begin{equation*}
		u_0(x) = \begin{cases}
		2,& \mbox{if}~ x < 0,\\
		-1,& \mbox{if}~ x > 0.
		\end{cases}
	\end{equation*} 
In the experiment we take $T=1$ and $[\xi_{\min}, \xi_{\max}] = [-1.05, 2.05]$.  

Figure~\ref{ex2-fig-Burgers} shows the measure characterized by $F(T,x,\xi)$, solution $u(T,x)$, energy $\hat{E}(T,x)$, the evolution of total energy $ \int_{\Omega}\hat{E} dx$ and the total energy defects $\int_{\Omega}\mathfrak{E} dx $ obtained with $(N_t,N_x,N_{\xi}) = (150,200,200)$.  Further, the $L^p$-errors and corresponding rates on consecutively refined meshes are presented in Table~\ref{table2}. 
\end{example}

\begin{table}[htbp]
	\centering
	\caption{Experiment 2 for the inviscid Burgers equation (shock wave): error and convergence rate.  } \label{table2}
	\begin{tabular}{|c|cccc|}
		\hline
		$(N_t,N_x,N_f)$ & $L^1$-error &  rate & $L^2$-error  & rate \\
		\hline
		(10, 15, 10) & 2.2612e-01&  -	&3.3309e-01&	- \\
		(20, 30, 20) & 9.5736e-02&  1.2400  	&2.3111e-01&	0.5273 \\
		(40, 60, 40) & 4.1352e-02&  1.2111 	&1.3917e-01&	0.7317 \\
		(80, 120, 80) & 2.4833e-02&  0.7357	&1.0081e-01&	0.4652 \\
		(160, 240, 160) & 9.7789e-03&   1.3445	&7.0210e-02&	0.5219 \\
		\hline
	\end{tabular}
\end{table}

\begin{figure}[htbp]
	\setlength{\abovecaptionskip}{0.cm}
	\setlength{\belowcaptionskip}{-0.cm}
	\centering

 \begin{subfigure}{0.5\textwidth}
        \centering
        \includegraphics[width=\linewidth]{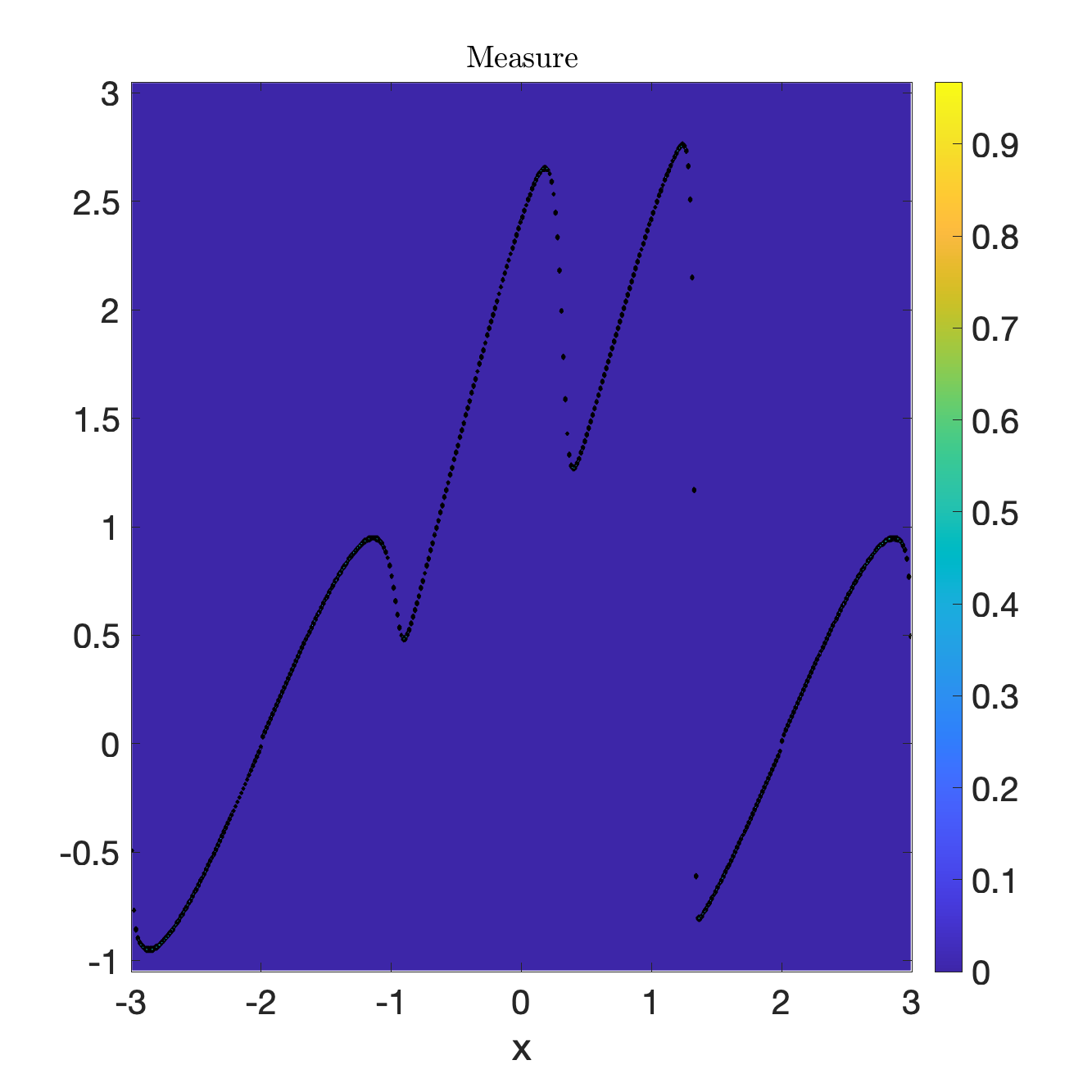}
    	\end{subfigure}
	\begin{subfigure}{0.48\textwidth}
        \centering
        \includegraphics[width=\textwidth]{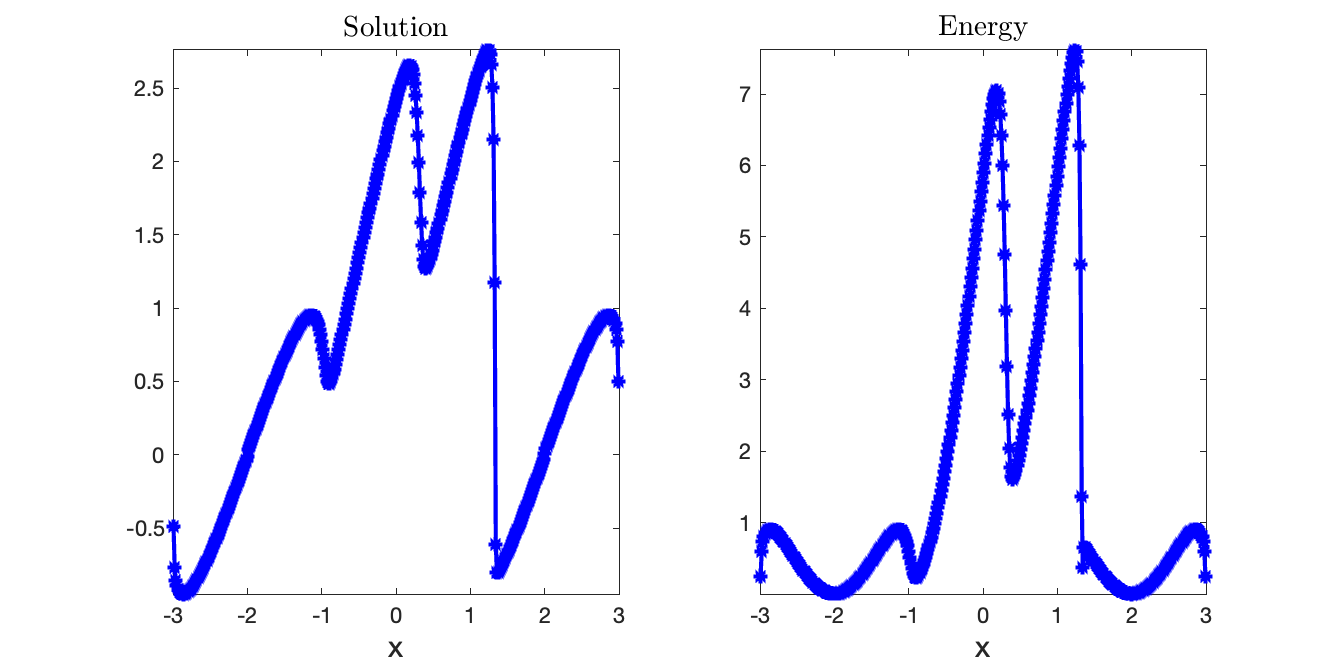}\\
       \includegraphics[width=\textwidth]{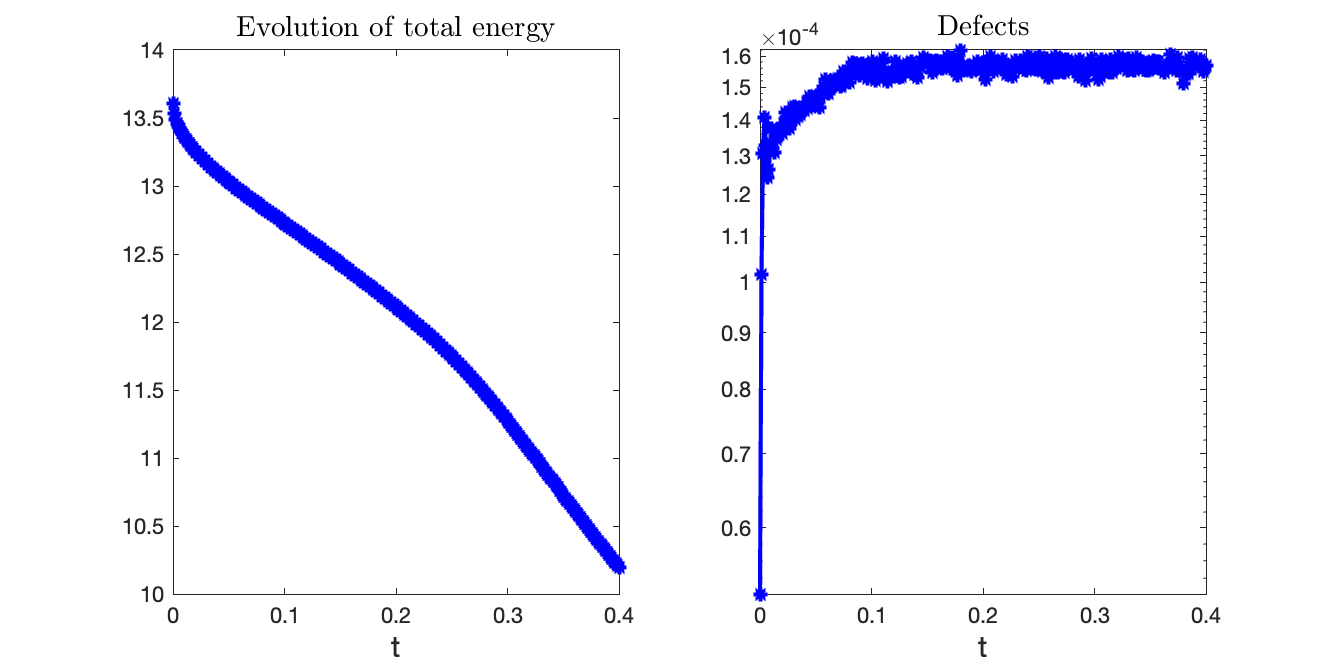}
    	\end{subfigure}
	\caption{  \small{Experiment 3 for the  inviscid Burgers equation:  the measure characterized by $F(T,x,\xi)$, solution $u(T,x)$, energy $\hat{E}(T,x)$, the evolution of total energy $ \int_{\Omega}\hat{E} dx$ and the total energy defects $\int_{\Omega}\mathfrak{E} dx $ with $T=0.4$. $(N_t,N_x,N_f) = (300,400,401)$.}}\label{ex3-fig-Burgers}
\end{figure}

\begin{example}\label{example:1D-compound-wave}\rm
The computational domain is $[-3,3]$ and the periodic boundary condition is applied.
The initial data are
\begin{equation*}
u_0(x) =  
\begin{cases}
 \sin (\pi x) & \mbox{if}~ |x| \geq 1, \\
3 & \mbox{if}~ x \in (-1,-0.5] \bigcup (0,0.5],\\
1 & \mbox{if}~ x \in (-0.5,0] ,\\
2 & \mbox{if}~ x \in (0.5,1].
\end{cases}
 \end{equation*} 
 In this experiment, we set $T=0.4$ and $[\xi_{\min}, \xi_{\max}] = [-1.05, 3.05]$.  

Figure~\ref{ex3-fig-Burgers} shows the measure characterized by $F(T,x,\xi)$, solution $u(T,x)$, energy $\hat{E}(T,x)$, the evolution of total energy $ \int_{\Omega}\hat{E} dx$ and the total energy defects $\int_{\Omega}\mathfrak{E} dx $ obtained with $(N_t,N_x,N_f) = (300,400,401)$.  
\end{example}

\subsection{The barotropic Euler equations} \label{sec:beuler}
We denote by $\hat{E}$ and $\mathfrak{E}$ 
the energy  and the energy defect, respectively,
\begin{align}
& \hat{E} = \int_{\mathbb{R}^2} \left( \frac{|\widetilde{\vm}|^2}{2\tilde{\vr}} + \frac{\tilde{\vr}^{\gamma}}{\gamma-1} \right) F(t, x, \tilde{\vr}, \widetilde{\vm}) d \tilde{\vr} d \widetilde{\vm}, \\
& \mathfrak{E} = \hat{E} - \left( \frac{|\vm|^2}{2\vr} + \frac{\vr^{\gamma}}{\gamma-1} \right).
\end{align}
Here, the solution (expected value with respect to the Young measure characterized by its probability distribution function) is
\[
(\vr, \vm) = \int_{\mathbb{R}^2}(\tilde{\vr}, \widetilde{\vm}) F(t, x, \tilde{\vr}, \widetilde{\vm}) d \tilde{\vr} d \widetilde{\vm}.
\]

We note that the reference solutions marked with black lines were obtained using the standard finite volume method with the Lax-Friedrichs numerical flux and $(N_t,N_x)=(2000,1000)$.  

\begin{figure}[htbp]
	\setlength{\abovecaptionskip}{0.cm}
	\setlength{\belowcaptionskip}{-0.cm}
	\centering

	\includegraphics[width=0.49\textwidth]{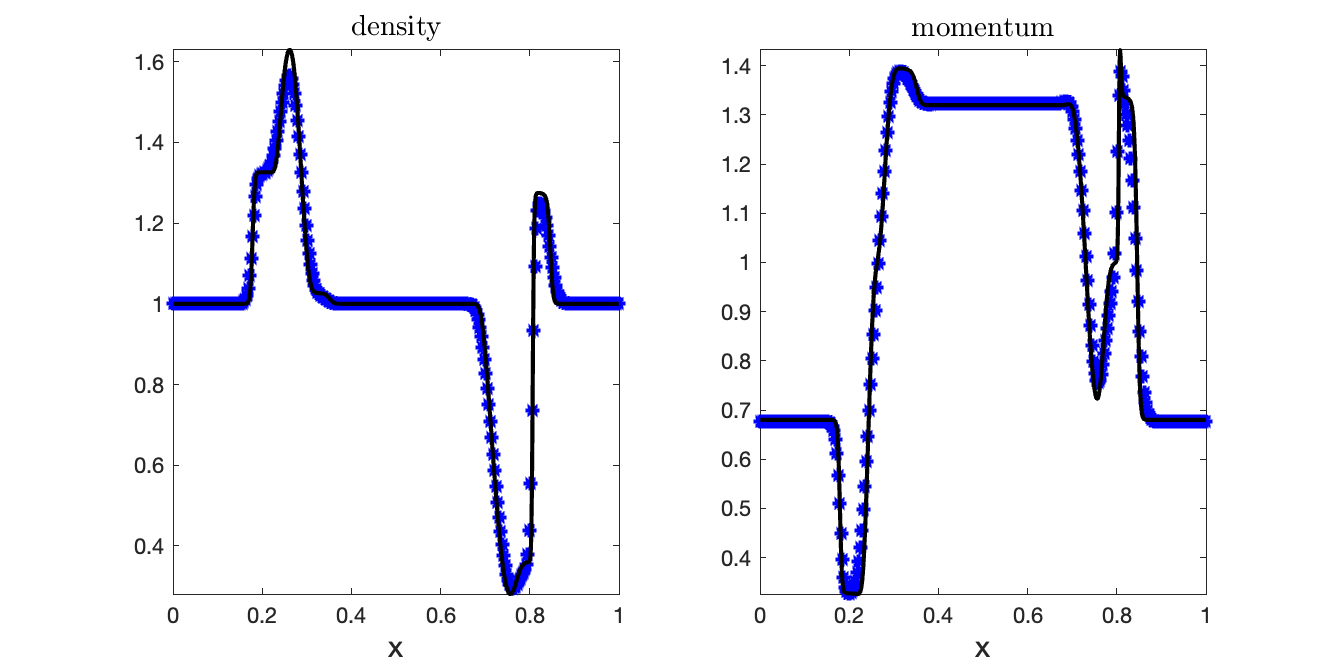} 
	\includegraphics[width=0.49\textwidth]{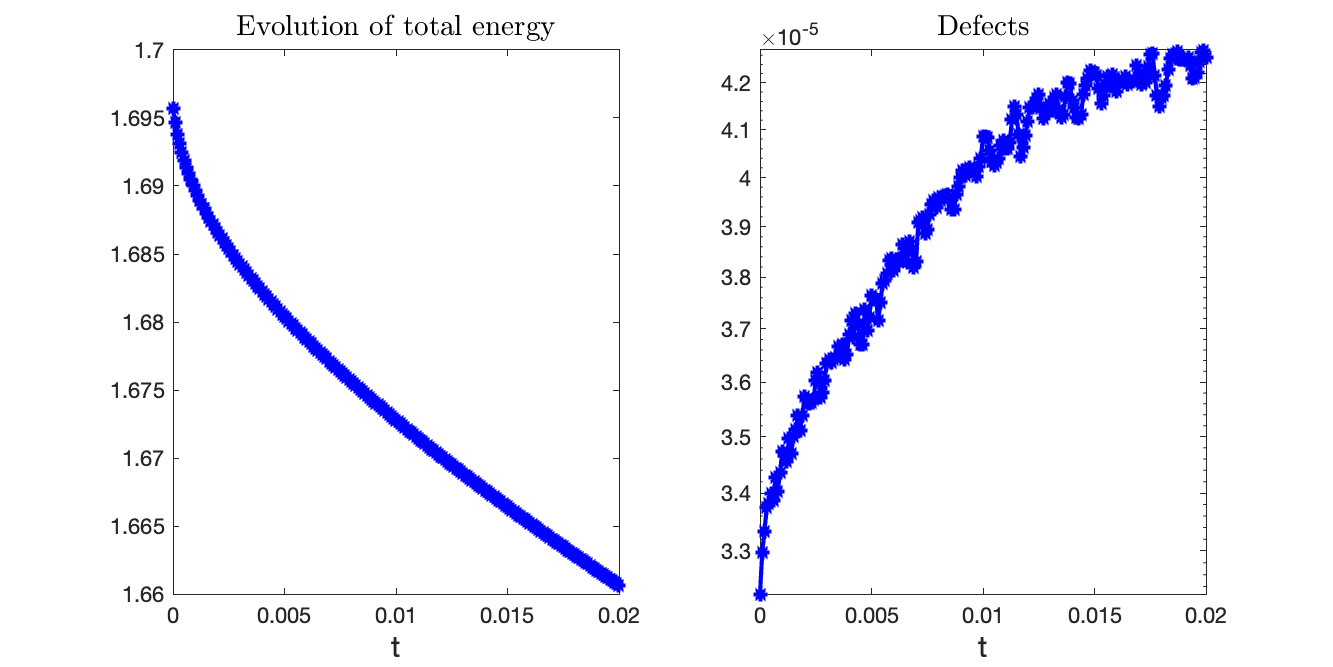}\\ 
	\caption{  \small{Experiment 1 for the barotropic  Euler system: density $\vr$, momentum $m$, the evolution of total energy $\int_{\Omega} \hat{E} dx$ and energy defect $\int_{\Omega} \mathfrak{E} dx$ obtained with $(N_t,N_x,N_{\vr},N_{m}) = (200,300,151,151)$.}}\label{ex1-fig-BEuler}
\end{figure}

\begin{example}[Degond-Tang problem \cite{DeTa:2011}]\rm

The computational domain is chosen to be $\Omega = [0,1]$ and the periodic boundary conditions are applied. The initial conditions read
\begin{align}
\left(\vr_0, \, m_0 \right)(x) = \begin{cases}
(1,\, 1-\epsilon^2/2) & \mbox{ if } x \in [0, 0.2] \cup [0.8,1], \\
(1+\epsilon^2,\, 1)	& \mbox{ if } x \in (0.2, 0.3], \\
(1,\, 1+\epsilon^2/2)	& \mbox{ if } x \in (0.3, 0.7], \\
(1-\epsilon^2,\, 1)	& \mbox{ if } x \in (0.7, 0.8].
\end{cases}
\end{align}
In this experiment, we take  $\epsilon = 0.8$, $T=0.06$ and $(\vr, m)\in [0.105, 1.805]\times[0.205,1.805]$.  

Figure~\ref{ex1-fig-BEuler} presents the density $\vr$, momentum $m$, the evolution of total energy $\int_{\Omega} \hat{E} dx$ and energy defect $\int_{\Omega} \mathfrak{E} dx$ obtained with  $(N_t,N_x,N_{\vr},N_{m}) = (200,300,151,151)$.  
\end{example}

\begin{figure}[htbp]
	\setlength{\abovecaptionskip}{0.cm}
	\setlength{\belowcaptionskip}{-0.cm}
	\centering
	
	\includegraphics[width=0.49\textwidth]{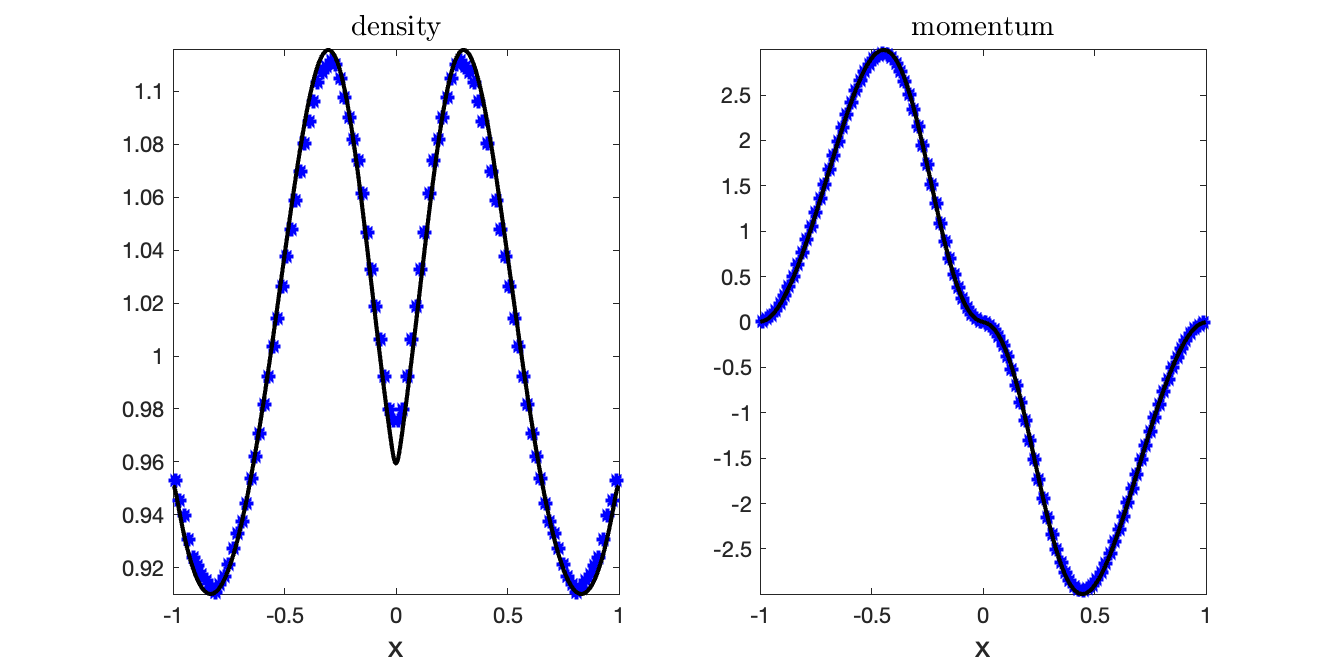} 
	\includegraphics[width=0.49\textwidth]{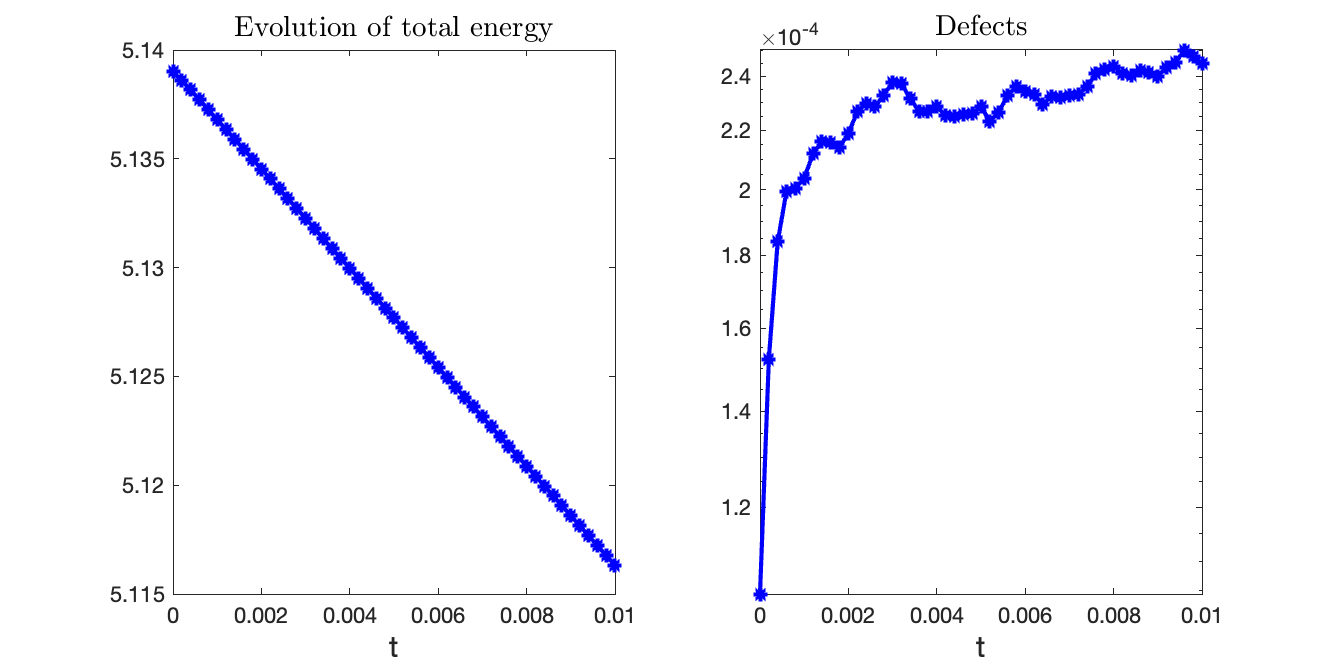}\\ 
	\caption{  \small{Experiment 2 for the Euler system: density $\vr$, momentum $m$, the evolution of total energy $\int_{\Omega} \hat{E} dx$ and energy defect $\int_{\Omega} \mathfrak{E} dx$ obtained with $(N_t,N_x,N_{\vr},N_{m}) = (50,100,51,201)$.}}\label{ex2-fig-BEuler}
\end{figure}

\begin{example}[Two colliding acoustic waves]\rm
The computational domain is set to $\Omega = [-1,1]$ and the periodic boundary conditions are applied. The initial conditions are given as
\begin{align}
\vr_0(x) = 0.955 + 0.5 \epsilon (1-\cos(2\pi x)), \quad
u_0(x) = -\mbox{sign}(x) \sqrt{2} (1-\cos(2\pi x)). 
\end{align}
In the experiment we set  $\epsilon = 0.1$, $T=0.01$ and $(\vr, m) \in [0.805, 1.205]\times[-3.105,3.105]$.  

Figure~\ref{ex2-fig-BEuler} shows the density $\vr$, momentum $m$, the evolution of total energy $\int_{\Omega} \hat{E} dx$ and energy defect $\int_{\Omega} \mathfrak{E} dx$ obtained with $(N_t,N_x,N_{\vr},N_{m}) = (50,100,51,201)$.  
\end{example}

\begin{figure}[htbp]
	\setlength{\abovecaptionskip}{0.cm}
	\setlength{\belowcaptionskip}{-0.cm}
	\centering

	\includegraphics[width=0.49\textwidth]{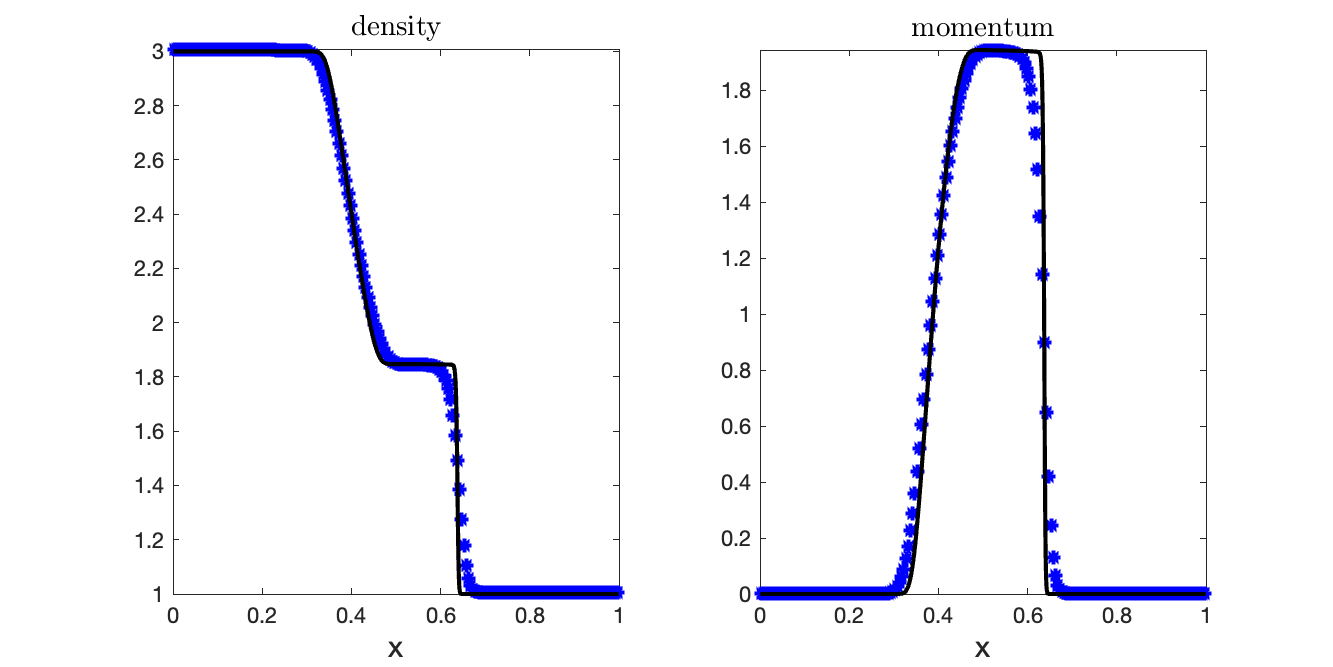} 
	\includegraphics[width=0.49\textwidth]{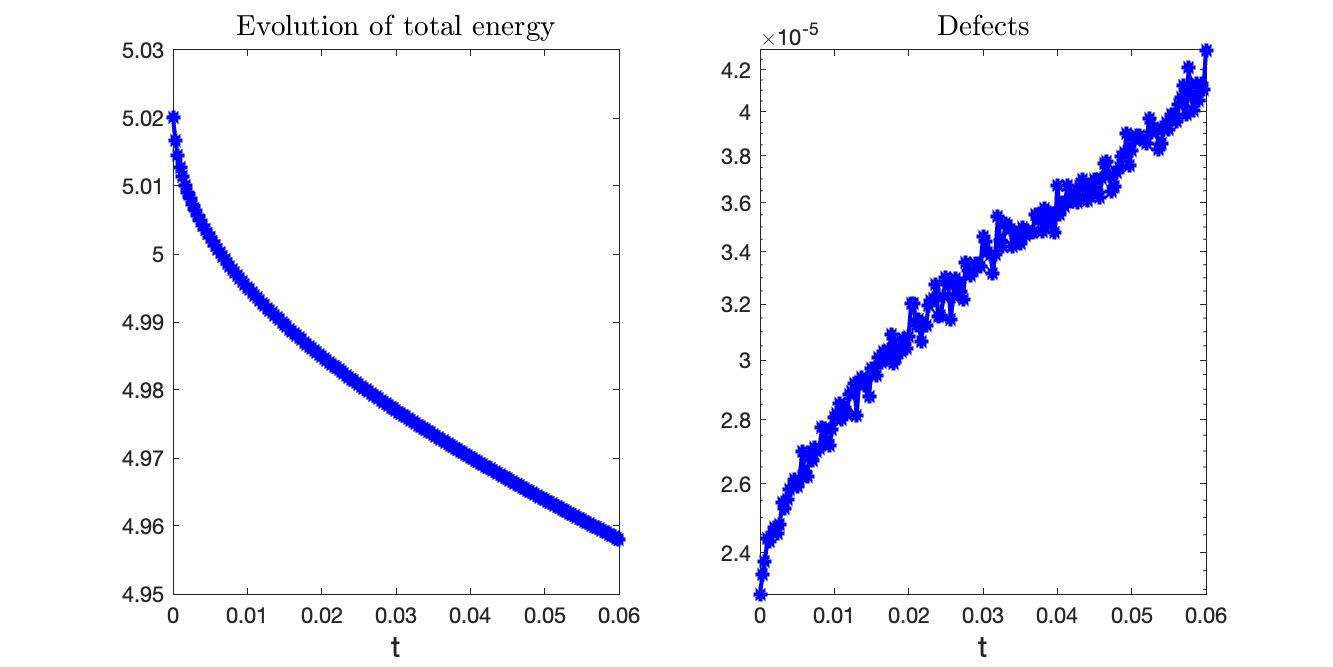}\\ 
	\caption{  \small{Experiment 3 for the barotropic Euler system: density $\vr$, momentum $m$, the evolution of total energy $\int_{\Omega} \hat{E} dx$ and energy defect $\int_{\Omega} \mathfrak{E} dx$ obtained with $(N_t,N_x,N_{\vr},N_{m}) =  (180,200,201,201)$}}\label{ex3-fig-BEuler}
\end{figure}

\begin{example}[Riemann problem]\rm
The computational domain is taken to  be $\Omega = [0,1]$ and the outflow boundary conditions are used. The initial conditions are chosen to be as follows
\begin{align}
\vr_0(x) = \begin{cases}
3 & \mbox{ if } x < 0.5, \\
1	& \mbox{ if } x \geq 0.5,
\end{cases} \quad
m_0(x) = 0. 
\end{align}
In this experiment, we set $T=0.06$ and $(\vr, m) \in [0.505, 3.505]\times[-0.505,2.005]$.  

Figure~\ref{ex3-fig-BEuler} shows density $\vr$, momentum $m$, the evolution of total energy $\int_{\Omega} \hat{E} dx$ and energy defect $\int_{\Omega} \mathfrak{E} dx$ obtained with $(N_t,N_x,N_{\vr},N_{m}) = (180,200,201,201)$.  
\end{example}

\subsection{The Allen-Cahn equation} \label{sec:allencahn}
Finally, we demonstrate the applicability of the LP formulation with the one-dimensional Allen-Cahn equation
\begin{align}\label{pde-AC}
\partial_t u  = \partial_{xx} u - G'(u), \quad G(u) = \frac{(1-u^2)^2}{4}, \quad \Omega = [0,1],
\end{align}
where periodic boundary conditions are applied.

Clearly, the potential energy $G(u)$ is double-well and not convex. Therefore, in general, minimizing $G(u)$ does not yield a Dirac measure, 
which is confirmed by numerical simulations for Experiment \ref{EX:AC-2}. 
The Young measure obtained by minimizing $G(u)$ is a convex combination of two Dirac measures sitting at $\xi = \pm 1$, whereas its corresponding solution, i.e., expected value with respect to the Young measure, agrees well with the reference solution. We do not present these numerical simulations here and leave them for interested readers. 

\medskip
As the one-dimensional Allen-Cahn equation \eqref{pde-AC} admits a unique (weak) solution for a large class of initial data (e.g.\ $u_0 \in L^2(\Omega)),$ our aim is to obtain a Dirac measure for the LP problem \eqref{LP1} - \eqref{LP3}. 
Inspired by the weak-strong uniqueness via a regularized energy functional \cite{BEHL}, here we employ the regularization $\tilde G$ of the 
potential energy $G$ yielding a convex cost functional in the resulting  LP problem
\begin{align*}
\widetilde{G}(u) = G(u) + \frac{\alpha}{2} u^2, \quad \alpha > 1.
\end{align*}
Consequently, the LP problem reads \begin{align}
\mbox{argmin}  \int_{\Omega}\int_{\mathbb{R}}   \widetilde{G}(\xi) F(t,x,\xi) ~d \xi  dx,\quad 
\end{align}
subject to
\begin{align}
& \int_{\mathbb{R}} F(t,x,\xi) d\xi=1,   \qquad  F(t, x, \xi) \geq 0, \\
&\int_{\mathbb{R}}\Big( \xi \partial_t F(t,x,\xi) -  \xi \Lap F(t,x,\xi) + G'(\xi) F(t,x,\xi) \Big) d\xi   = 0.
\end{align}
Accordingly, we again apply a finite volume method to obtain a finite-dimensional discrete formulation:
\begin{align}
   &  \text{argmin}_F \sum_{k=1}^{N_x}\sum_{l=1}^{N_{\xi}} \Ov{\widetilde{G}}_l F_{kl}^j 
\end{align}
subject to 
\begin{align}
& \sum_{l} F_{kl}^j =1, \qquad 
    F_{jk}^l \geq 0, \\  
& \sum_{l} \left(\Ov{u}_l \frac{F_{kl}^{j+1}-F_{kl}^j}{\Delta t} - \Ov{u}_l \frac{F_{k+1,l}^j- 2 F_{k,l}^j +F_{k-1,l}^j}{h_x^2} + \Ov{G'}_l F_{k,l}^j  \right)=0,  
\end{align}
with
\begin{align}\label{not}
\Ov{u}_l = \frac{1}{h_{\xi}}\int_{K_{\xi}^l} \xi d \xi, \quad  
\Ov{\widetilde{G}}_l = \frac{1}{h_{\xi}}\int_{K_{\xi}^l} \widetilde{G}(\xi) d \xi,  \quad 
\Ov{G'}_l = \frac{1}{h_{\xi}}\int_{K_{\xi}^l} G'(\xi) d \xi.
\end{align}

Let us denote the solution (expected value with respect to the Young measure) as follows
\[
u = \int_{\mathbb{R}} \xi F(t, x, \xi) d \xi.
\] 
Further, we define the energies 
\begin{align*}
E = G(u) + \frac12 |\Grad u|^2, \quad 
\hat{E} = \int_{\mathbb{R}} G(\xi) F(t, x, \xi) d \xi + \frac12 \int_{\mathbb{R}} \left( \Grad F(t, x, \xi)  \right)^2 d \xi, \\
E^{RE} = \widetilde{G}(u) + \frac12 |\Grad u|^2, \quad 
\hat{E}^{RE} = \int_{\mathbb{R}} \widetilde{G}(\xi) F(t, x, \xi) d \xi + \frac12 \int_{\mathbb{R}} \left( \Grad F(t, x, \xi)  \right)^2 d \xi,
\end{align*}
as well as the corresponding defects
\begin{align*}
\mathfrak{E} = \hat{E} - E , \quad \mathfrak{E}^{RE} = \hat{E}^{RE} - E^{RE}. 
\end{align*}
In the following simulations, we set $\alpha = 1.1$ and use the central difference to approximate $\Grad$. Moreover, the reference solutions marked with black lines are obtained by a standard finite volume method with $(N_t,N_x)=(100000,1000)$.

\begin{figure}[htbp]
	\setlength{\abovecaptionskip}{0.cm}
	\setlength{\belowcaptionskip}{-0.cm}
	\centering
	\includegraphics[width=0.48\linewidth]{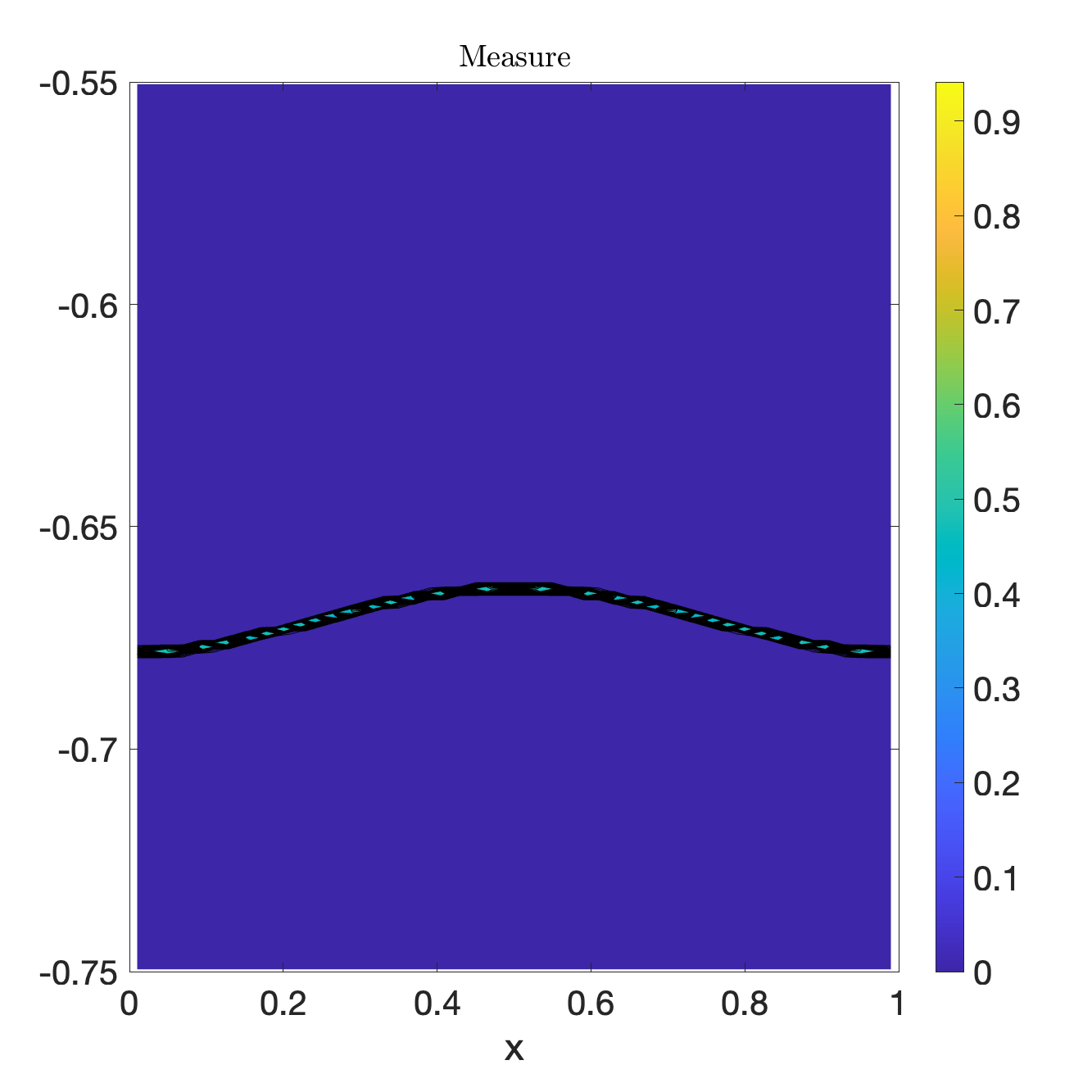}
         \includegraphics[width=0.5\textwidth]{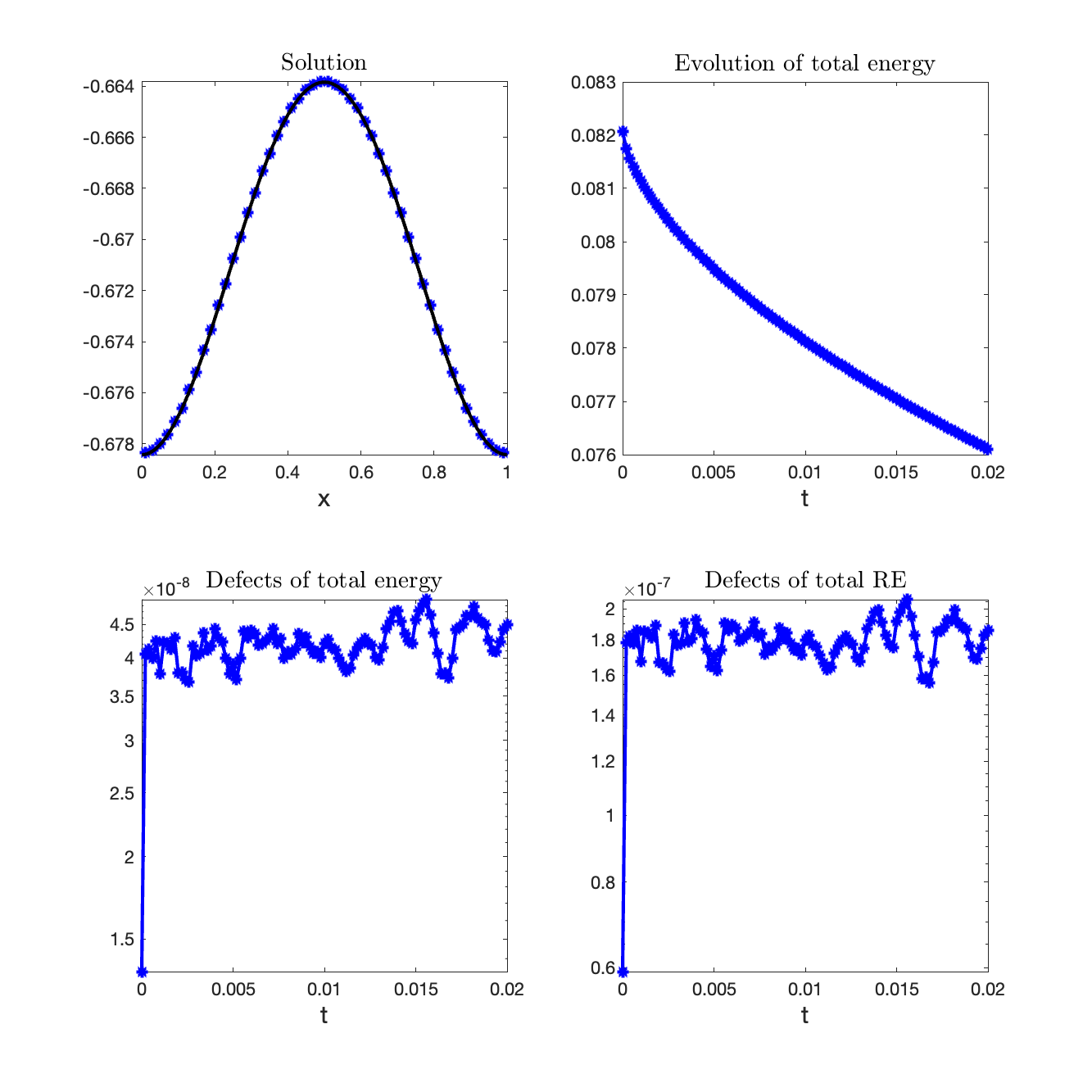}\\
	
	\caption{  \small{Experiment 1 for the  Allen-Cahn equation (Double interfaces): the measure characterized by $F(T,x,\xi)$, solution $u(T,x)$, 
the evolution of total energy $ \int_{\Omega}\hat{E} dx$,  the defect of total energy $\int_{\Omega}\mathfrak{E} dx$ as well as the defect of total regularized energy $\int_{\Omega}\mathfrak{E}^{RE} dx$ obtained with $(N_t,N_x,N_{U}) = (100,50,200)$.}}\label{ex2-fig-AC}
\end{figure}

\begin{example}[Double interfaces]\rm 
The computational domain is taken to be $\Omega = [0,1]$. The initial data read
\begin{align}
u_0(x) = \tanh \left( \frac{ x - 0.25}{\sqrt{2}}\right) -  \tanh \left( \frac{ x - 0.75}{\sqrt{2}}\right) -1.
\end{align} 
We take $T = 0.02$ and $[\xi_{\min}, \xi_{\max}] = [-0.75, -0.55]$.

Figure~\ref{ex2-fig-AC} shows the measure characterized by $F(T,x,\xi)$, solution $u(T,x)$, 
the evolution of total energy $ \int_{\Omega}\hat{E} dx$,  the defect of total energy $\int_{\Omega}\mathfrak{E} dx$ as well as the defect of total regularized energy $\int_{\Omega}\mathfrak{E}^{RE} dx$ obtained with $(N_t,N_x,N_{U}) = (100,50,200)$. 

\end{example}

\begin{figure}[htbp]
	\setlength{\abovecaptionskip}{0.cm}
	\setlength{\belowcaptionskip}{-0.cm}
	\centering
	\includegraphics[width=0.48\linewidth]{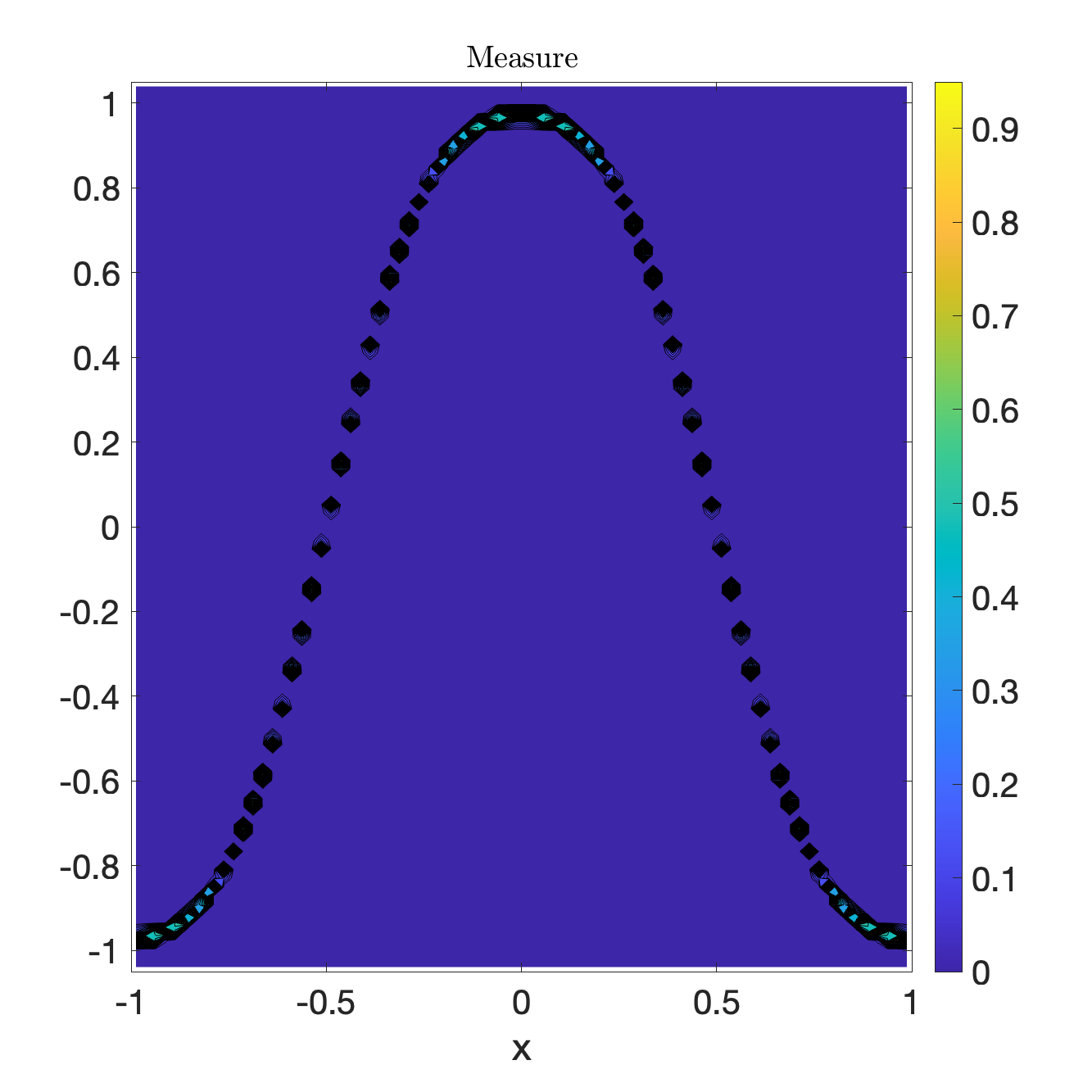}
    \includegraphics[width=0.5\textwidth]{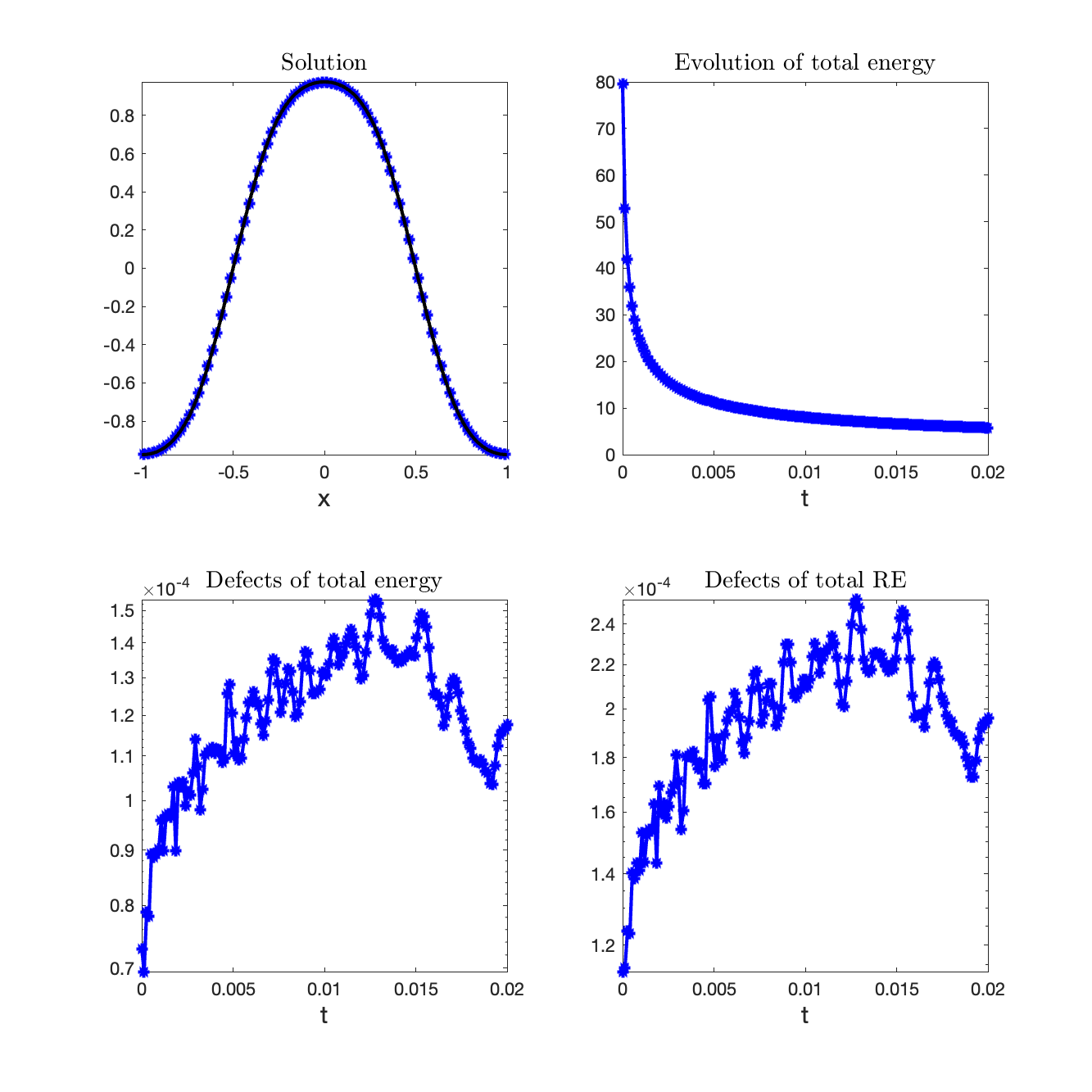}\\
	\caption{  \small{Experiment 2 for the  Allen-Cahn equation: the measure characterized by $F(T,x,\xi)$, solution $u(T,x)$, 
the evolution of total energy $ \int_{\Omega}\hat{E} dx$,  the defect of total energy $\int_{\Omega}\mathfrak{E} dx$ as well as the defect of total regularized energy $\int_{\Omega}\mathfrak{E}^{RE} dx$ obtained with $(N_t,N_x,N_{U}) = (150,80,100)$.}}\label{ex3-fig-AC}
\end{figure}

\begin{example}\label{EX:AC-2}\rm 
The computational domain is set to  $\Omega = [-1,1]$ and the initial data are chosen as follows
\begin{align}
u_0(x) = \begin{cases}
1, & \mbox{ if } |x| < 0.5, \\
-1, & \mbox{ else.}  
\end{cases}
\end{align}
In this experiment we take $T = 0.02$ and $[\xi_{\min}, \xi_{\max}] =  [-1.05, 1.05]$.  

\medskip 
Figure~\ref{ex3-fig-AC} presents the measure characterized by $F(T,x,\xi)$, solution $u(T,x)$, 
the evolution of total energy $ \int_{\Omega}\hat{E} dx$,  the defect of total energy $\int_{\Omega}\mathfrak{E} dx$, as well as the defect of total regularized energy $\int_{\Omega}\mathfrak{E}^{RE} dx$ obtained with $(N_t,N_x,N_{U}) = (150,80,100)$. 
\end{example}

\section{The Interior point method for LP} \label{app:1}
To solve for $\mathbf{F}$ using the interior point method on the primal problem, we have the following primal LP formulation:
\begin{align}
    & \text{argmin}_{\mathbf{F}} \Xi^T \mathbf{F}, \nonumber \\
     & \mathbf{F} \geq 0 \nonumber \\
        & M\mathbf{F}=\mathbf{c}, \quad M=\begin{pmatrix}
            A \\
            B
        \end{pmatrix},
\end{align}
where $M$ is a $2N_tN_x^d \times N_tN_x^dN^n_{\xi}$ matrix and $\mathbf{c}$ is a vector of size $2N_tN_x^d$. \\

To solve this problem, one can write the corresponding Lagrangian 
\begin{align}
    \mathcal{L}(\mathbf{F}, \Theta, \mathbf{s})=\Xi^T\mathbf{F}-\Theta^T(M\mathbf{F}-\mathbf{c})-\mathbf{s}^T\mathbf{F}. 
\end{align}
 We do not directly use gradient-descent methods on this Lagrangian, since that would be a first-order method. To guarantee optimality, we need to satisfy the KKT conditions. The first of such conditions comes from 
 \begin{align}
     \nabla_{\mathbf{F}} \mathcal{L}=\Xi-M^T\Theta-\mathbf{s}=0,
 \end{align}
 which is also the feasibility condition in the dual problem. The primal and dual problems are thus: 
\begin{enumerate}
    \item Primal: \begin{align}
         & \text{argmin}_{\mathbf{F}} \Xi^T \mathbf{F}, \nonumber \\
     & \mathbf{F} \geq 0 \nonumber \\
        & M \mathbf{F}=\mathbf{c}
    \end{align}
    where $\mathbf{F}$ has size $N_tN_x^d N_{\xi}$ and $\mathbf{c}$ has size $2N_tN_x^d$. 
    \item Dual: \begin{align}
         & \text{argmax}_{\Theta} \Theta^T \mathbf{c}, \nonumber \\
     & \mathbf{s} \geq 0 \nonumber \\
        & \Xi-M^T\Theta-\mathbf{s}=0 
    \end{align}
    where $\Theta$ has size $N_tN_x^d$ (the Lagrange multiplier to the equality constraint) and $\mathbf{s}$ (the Lagrange multiplier to the $\mathbf{F}\geq 0$ inequality constraint) has size $N_tN_x^dN_{\xi}$.
\end{enumerate}
The KKT conditions for optimality (i.e. $(\mathbf{F}^*, \Theta^*, \mathbf{s}^*)$ are optimal if and only if) are:  
\begin{enumerate}
    \item Primal feasibility: $M\mathbf{F}= \mathbf{c}$, $\mathbf{F}\geq 0$, 
    \item Dual feasibility: $M^T \Theta+\mathbf{s}=\Xi$, $\mathbf{s} \geq 0$
    \item Complementarity: $\mathbf{F}_{i} \mathbf{s}_i=0, \forall i$
\end{enumerate}
To use interior point methods, we replace the complementarity condition with a perturbed complementarity condition 
\begin{align}
    \mathbf{F}_{i}\mathbf{s}_i \rightarrow FS\mathbf{e}=\tau \mathbf{e}, \qquad \tau>0
\end{align}
where $F=\text{diag}(\mathbf{F})$, $S=\text{diag}(\mathbf{s})$, $\mathbf{e}=(1, \cdots, 1)^T$. With the replacement of the perturbed complementarity condition, this leads to the log-barrier 
\begin{align}
    \mathcal{L}(\mathbf{F}(\tau), \Theta)=\Xi^T \mathbf{F}(\tau)-\Theta^T(M\mathbf{F}(\tau)-\mathbf{c})-\tau \log(-\mathbf{F}(\tau)),  
\end{align}
but we also do not directly use gradient descent on this Lagrangian since it's still a first-order method and the problem becomes ill-conditioned as $\tau$ becomes small.\\

Instead, we look at the perturbed KKT system 
\begin{align}
    G_{\tau}(\mathbf{F}, \Theta, \mathbf{s})=\begin{pmatrix}
        M\mathbf{F}-\mathbf{c}\\
        M^T \Theta+\mathbf{s}-\Xi\\
        FS\mathbf{e}-\tau \mathbf{e}
    \end{pmatrix}=0,
\end{align}
where the solutions trace a central path and strict positivity $\mathbf{F}\geq 0, \mathbf{s} \geq 0$ are enforced, and as $\tau \rightarrow 0$, $(\mathbf{F}(\tau), \Theta(\tau), \mathbf{s}(\tau)) \rightarrow (\mathbf{F}^*, \Theta^*, \mathbf{s}^*)$.\\

To solve this with the interior point method, we use the primal-dual Newton step (which is a core step in the interior point method). Here we want $G_{\tau} \rightarrow 0$, but we cannot reach it directly; instead, we approach the solution using Newton's method to identify where the zero occurs. So we define the following residuals
\begin{align}
    r_p= M\mathbf{F}-\mathbf{c}, \quad r_d= M^T \Theta+\mathbf{s}-\Xi, \quad r_c= FS\mathbf{e}-\tau \mathbf{e}. 
\end{align}
Using Newton's method for finding the zeros of $G_{\tau}(z)$, it's clear that each Newton step $\Delta z$ has size
\begin{align}
   G_{\tau}(z+\Delta z) \rightarrow 0 \implies G'_{\tau}(z) \Delta z=-G_{\tau}(z),
\end{align}
and we can write it out explicitly 
\begin{align} 
   & G_{\tau}(z+\Delta z)=\begin{pmatrix}
         M(\mathbf{F}+\Delta \mathbf{F})-\mathbf{c}\\
        M^T (\Theta+\Delta \Theta)+\mathbf{s}+\Delta \mathbf{s}-\Xi\\
        (F+\Delta F)(S+\Delta S)\mathbf{e}-\tau \mathbf{e}
    \end{pmatrix} \approx \begin{pmatrix}
        r_p+M \Delta \mathbf{F} \\
        r_d+M^T \Delta \Theta+\Delta \mathbf{s} \\
        r_c+F\Delta S \mathbf{e}+\Delta F S \mathbf{e}  
    \end{pmatrix}\rightarrow 0 \implies \nonumber 
    \end{align}

\begin{align} \label{eq:invert}
     \begin{pmatrix} 
        0 & M^T & I \\
        M & 0 & 0 \\
        S & 0 & F
    \end{pmatrix} \begin{pmatrix}
        \Delta \mathbf{F} \\
        \Delta \Theta \\
        \Delta \mathbf{s}
    \end{pmatrix}=
    -\begin{pmatrix}
        r_d \\
        r_p \\
        r_c
    \end{pmatrix}.
\end{align}
After solving the above system at each step, $\tau$ is reduced, and the process is repeated until $\tau$ is sufficiently small and we obtain a good approximation of the desired optimal solution. \\

A common way of dealing with the above linear system is to reduce it using the Schur complement. 
One thus updates $\mathbf{F}$ and $\mathbf{s}$ to ensure positivity. Finally, to update $\tau$, one can choose 
\begin{align}
    \tau=\mu \frac{\mathbf{F}^T \mathbf{s}}{N_tN_x^dN_{\xi}}, \quad \mu \in (0, 1)
\end{align}
and repeat the process until $\mu$, $\|r_d\|$, $\|r_p\|$ are small. \\

We can summarise the classical cost in this procedure with $s$ variables and $r$ constraints: 
\begin{enumerate}
    \item Cost per iteration: We have a sparse matrix $M$, then we can use sparse factorization 
    \item Number of iterations: $\sqrt{s}\log(1/\epsilon)$ 
\end{enumerate}
The earliest quantum algorithms based on primal-dual interior point methods are the preparation of quantum states $|\Delta F\rangle$, $|\Delta s\rangle$, and $|\Theta\rangle$ at each Newton iteration step \cite{kerenidis2020quantum}, then performing tomography at each step before going to the next step. The cost of using quantum linear systems of equations (QLSA) to invert the matrix in Eq.~\eqref{eq:invert} depends on the condition number $\kappa_{Newt}$ of this matrix. 
 \end{document}